\documentclass[a4paper,11pt]{article}
\usepackage{jinstpub} 

\title{\boldmath Enhancing Particle Identification in Helium-Based Drift Chambers Using Cluster Counting: Insights from Beam Test Studies}

\author[a,m,1]{W. Elmetenawee \note{Corresponding author.}}
\author[a,b]{M.~Abbrescia}
\author[a,d]{M.~Anwar}
\author[n]{C.~Caputo}
\author[l]{G.~Chiarello}
\author[c]{A.~Corvaglia}
\author[a]{F.~Cuna}
\author[a,b]{B.~D'Anzi}
\author[a,d]{N.~De~Filippis}
\author[c,e]{F.~De~Santis}
\author[g,h]{M.~Dong}
\author[c,e]{E.~Gorini}
\author[c]{F.~Grancagnolo}
\author[c,e]{S.~Grancagnolo}
\author[c,e]{F.G.~Gravili}
\author[j]{M.~Greco}
\author[f]{K.F.~Johnson}
\author[g]{S.~Liu}
\author[a,b]{M.~Louka}
\author[n]{P.~Mastrapasqua}
\author[c]{A.~Miccoli}
\author[c,e]{M.~Panareo}
\author[c]{M.~Primavera}
\author[a,d]{F.M.~Procacci} 
\author[i]{A.~Taliercio} 
\author[c,k]{G.F.~Tassielli}
\author[c,e]{A.~Ventura}
\author[g]{L.~Wu}
\author[g]{G.~Zhao}

\affiliation[a]{Istituto Nazionale di Fisica Nucleare Sezione di Bari, Via E. Orabona 4, 70126 Bari , Italy}
\affiliation[b]{Universitá di Bari "Aldo Moro", Via E. Orabona 4, 70126 Bari, Italy}
\affiliation[c]{Istituto Nazionale di Fisica Nucleare Sezione di Lecce, Via Arnesano, 73100 Lecce, Italy}
\affiliation[d]{Politecnico di Bari, Via Amendola 126/b, 70126 Bari, Italy}
\affiliation[e]{Universitá del Salento, Via Arnesano, 73100 Lecce, Italy}
\affiliation[f]{Florida State University, 600 W College Ave, Tallahassee FL, 32306, United States}
\affiliation[g]{Institute of High Energy Physics, Chinese Academy of Sciences, Beijing 100049, China}
\affiliation[h]{University of Chinese Academy of Sciences, Beijing 100049, China}
\affiliation[i]{Northwestern University, 2025 Campus Dr, Evanston, Illinois 60208, United States}
\affiliation[j] {Max-Planck-Institut für Physik, Boltzmannstr. 8, 85748 Garching, Germany}
\affiliation[k] {Universitas Mercatorum, Piazza Mattei 10, Roma, RM 00186, Italy}
\affiliation[l] {Department of Engineering, University of Palermo, Viale delle Scienze 9, Palermo, 90128, Italy}
\affiliation[m] {Physics Department, Faculty of Science, Helwan University, Cairo, 11792 Helwan, Egypt}
\affiliation[n] {Universite Catholique de Louvain (UCL), Pl. de l'Université 1, 1348 Ottignies-Louvain-la-Neuve, Belgium}
\emailAdd{walaa.elmetenawee@ba.infn.it}

\abstract{Particle identification in gaseous detectors traditionally relies on energy loss measurements (dE/dx); however, uncertainties in total energy deposition limit its resolution. The cluster counting technique (dN/dx) offers an alternative approach by exploiting the Poisson-distributed nature of primary ionization, providing a statistically robust method for mass determination. Simulation studies with Garfield++ and Geant4 indicate that dN/dx can achieve twice the resolution of dE/dx in helium-based drift chambers. However, experimental implementation is challenging due to signal overlap in the time domain, complicating the identification of electron peaks and ionization clusters. This paper presents novel algorithms and modern computational techniques to address these challenges, facilitating accurate cluster recognition in experimental data. The effectiveness of these algorithms is validated through four beam tests conducted at CERN, utilizing various helium gas mixtures, gas gains, and wire orientations relative to ionizing tracks. The experiments employ a muon beam (1 GeV/c–180 GeV/c) with drift tubes of different sizes and sense wire diameters. The analysis explores the Poisson nature of cluster formation, evaluates the performance of different clustering algorithms, and examines the dependence of counting efficiency on the beam particle impact parameter. Furthermore, a comparative study of the resolution achieved using dN/dx and dE/dx is presented.}


\arxivnumber{1234} 

\begin{document}
\maketitle
\flushbottom

\section{Introduction}
\label{sec:intro}
In high-energy physics colliders, particle identification (PID) in gaseous detectors has traditionally relied on measuring ionization energy loss (dE/dx) from charged particle tracks. While this method has been widely used, its effectiveness is constrained by significant uncertainties in total energy deposition. Even in the most favorable momentum region—where relativistic rise enhances separation—the overlap between energy loss distributions of different particle species often reduces the achievable resolution, limiting precise identification~\cite{CC}.

To overcome these limitations, the cluster counting technique (dN/dx) has been introduced as a more statistically robust alternative. This method exploits the Poisson-distributed nature of primary ionization, offering improved mass determination capabilities~\cite{CC, PID}. Unlike dE/dx, dN/dx is largely independent of fluctuations in cluster size and gas gain, and it remains unaffected by highly ionizing $\gamma$-rays, making it a more stable and reliable approach for PID.

The concepts of cluster counting and cluster timing have been pioneered in earlier drift chamber developments and are now being refined and integrated into modern detector designs. These techniques are anticipated to play a pivotal role in the drift chambers envisioned for next-generation collider experiments such as FCC-ee and CEPC~\cite{FCC, Elmetenawee}.

\section{The cluster counting technique: dN/dx}

In helium-based gas mixtures, ionization events in a gas detector produce signals that spread over a few nanoseconds. The use of fast readout electronics enables efficient signal identification, facilitating precise event reconstruction. By counting the number of ionization clusters per unit length (dN/dx), particle identification can be achieved with significantly higher resolution than conventional methods based on the total energy loss integral (dE/dx).

\begin{figure}[ht]
\centering
\includegraphics[width=0.8\textwidth,height=0.35\textheight] {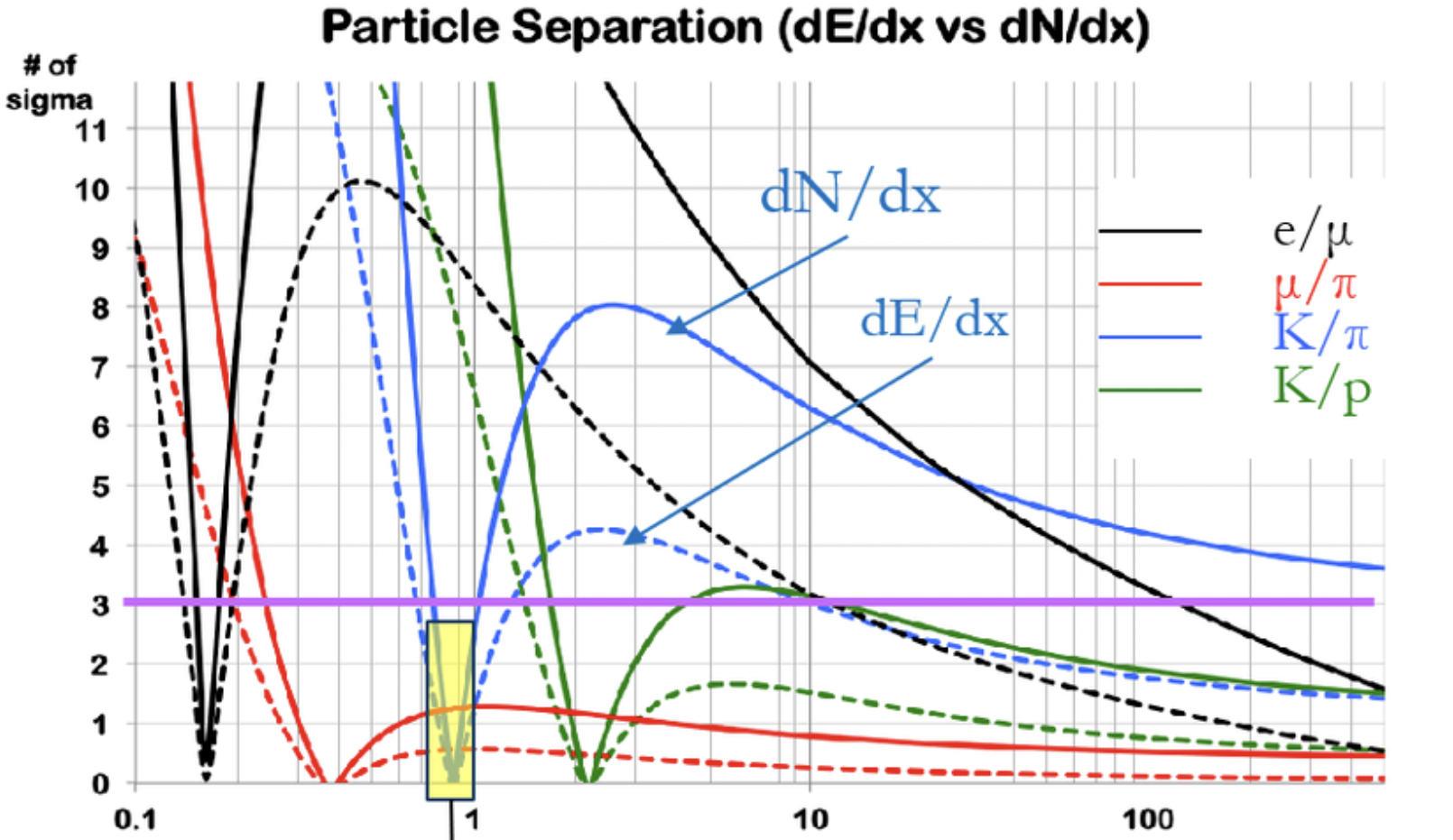}
\includegraphics[width=.6\textwidth,height=0.35\textheight] {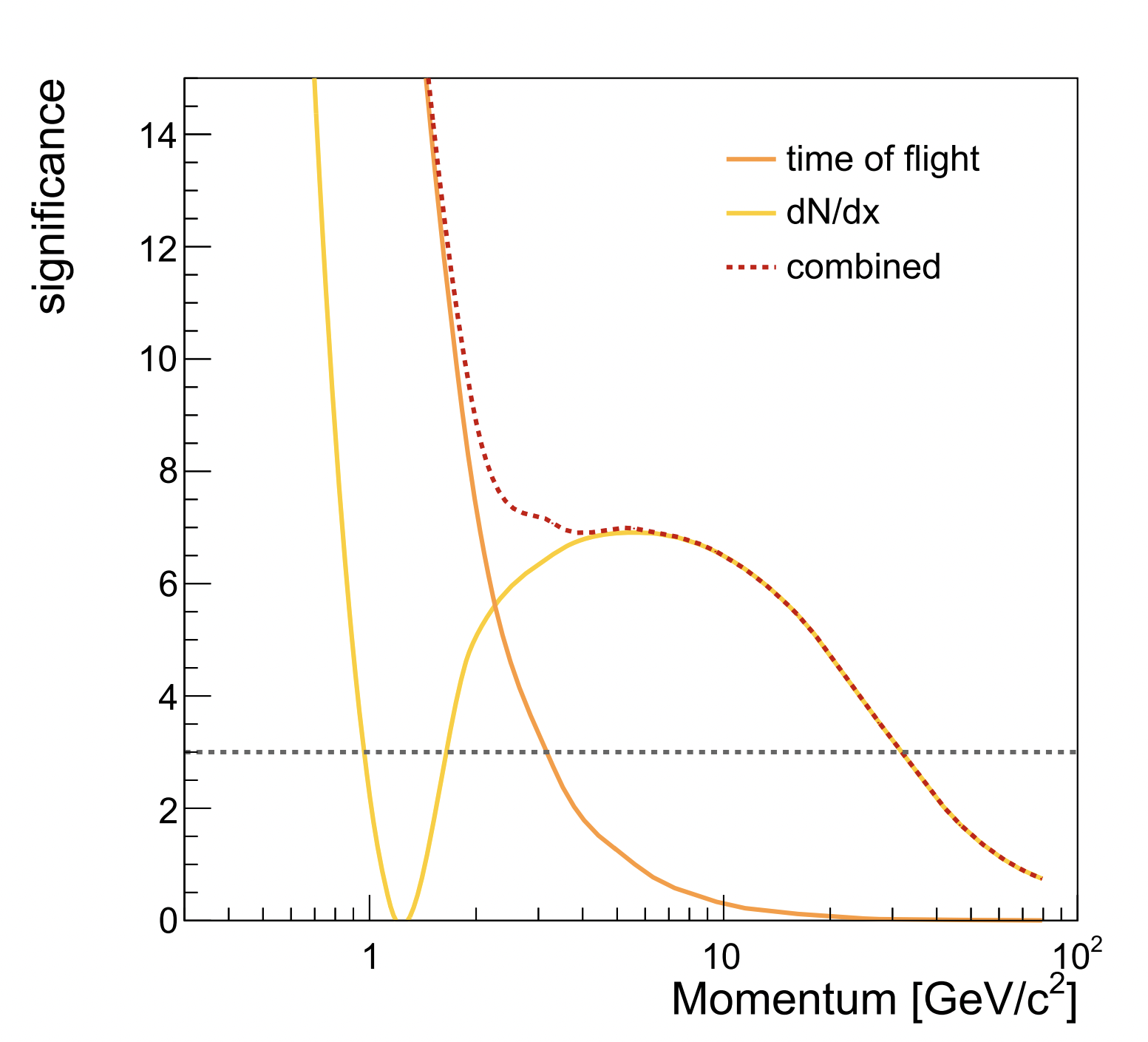}
\qquad \qquad
\caption{top: Analytic evaluation of particle separation capabilities achievable with dE/dx (dashed curves) and dN/dx (solid curves). The region between 0.85 GeV/c and 1.05 GeV/c where a different technique is needed, is highlighted in yellow. bottom: $K/\pi$ separation in number of $\sigma$ as a function of the particle momentum using the dN/dx method, the time-of-flight method and their combination.~\cite{Bedeschi}.}
\label{Fig:Analy}
\end{figure}

Analytical evaluations based on a parameterization of the Bethe–Bloch formula~\cite{Walenta}, shown in Figure~\ref{Fig:Analy} (top), predict that the cluster counting technique enhances particle separation by a factor of two compared to dE/dx. The plot presents the separation power, expressed in standard deviations ($\sigma$), as a function of momentum for a gas mixture of 90\% He and 10\% iC$_{4}$H$_{10}$. Solid curves correspond to the cluster counting technique, assuming a cluster counting efficiency of 80\%, 
 while dashed curves represent the optimal energy loss truncated mean method. The results highlight the robustness of the technique across a broad momentum range, with only minor performance variations in specific intervals.

Additionally, Figure~\ref{Fig:Analy} (bottom) displays the combined performance of the cluster counting and time-of-flight techniques, as obtained from the DELPHES fast simulation, as performed in the case of the IDEA proposed detector at FCC-ee~\cite{Bedeschi}. The results indicate efficient $K/\pi$ separation ($\geq 3\sigma$) for momenta p < 30 GeV, further demonstrating the effectiveness of this approach.

A dedicated simulation study has been conducted to evaluate the performance of the cluster counting technique and investigate ionization processes in a helium-based drift chamber, using both Garfield++ and Geant4~\cite{SIM}. The results confirm that cluster counting significantly improves particle separation compared to the conventional dE/dx method. However, slight discrepancies are observed between the separation power predicted by Geant4 and Garfield++. Moreover, experimental data on cluster density and population in helium-based gases remain scarce. Consequently, experimental validation is essential to confirm the effectiveness of the cluster counting technique and to benchmark the simulation results.

\section{Experimental Setup, Readout System, and Beam Test Campaigns}

\begin{figure}[ht]
\centering
\includegraphics[width=0.95\textwidth] {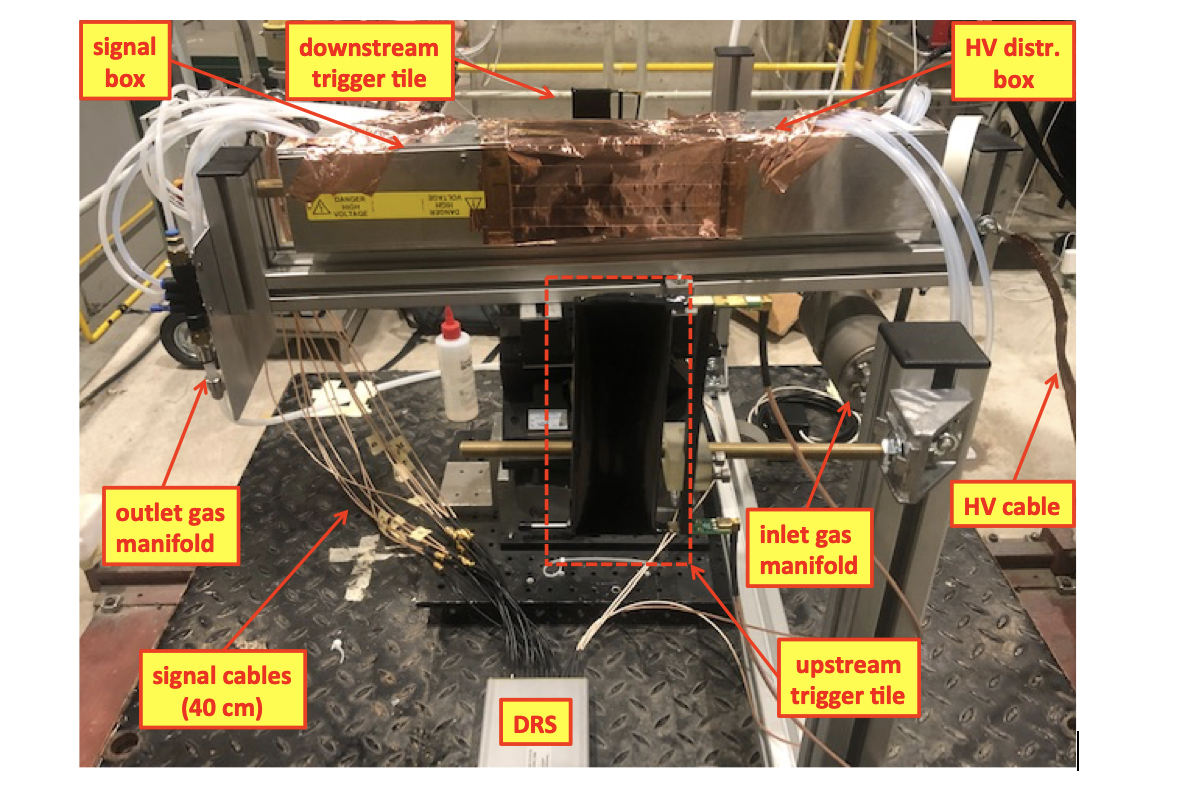}
\qquad 
\caption{The experimental setup,  seen from the beam upstream, for a beam test performed in 2021. The main components of the setup are indicated by the yellow insets.}
\label{Fig:Setup}
\end{figure}

\begin{figure}[ht]
\centering
\includegraphics[width=0.45\textwidth] {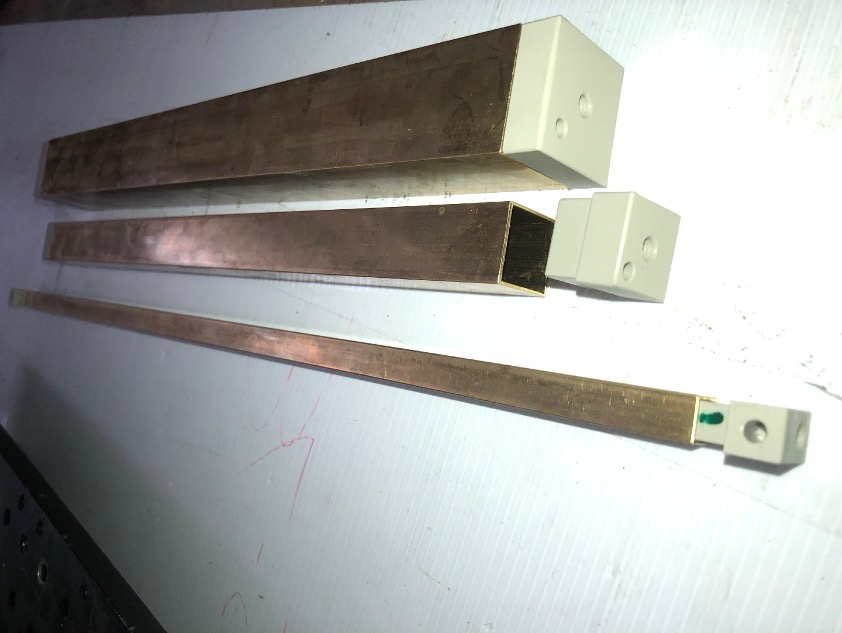}
\includegraphics[width=0.45\textwidth] {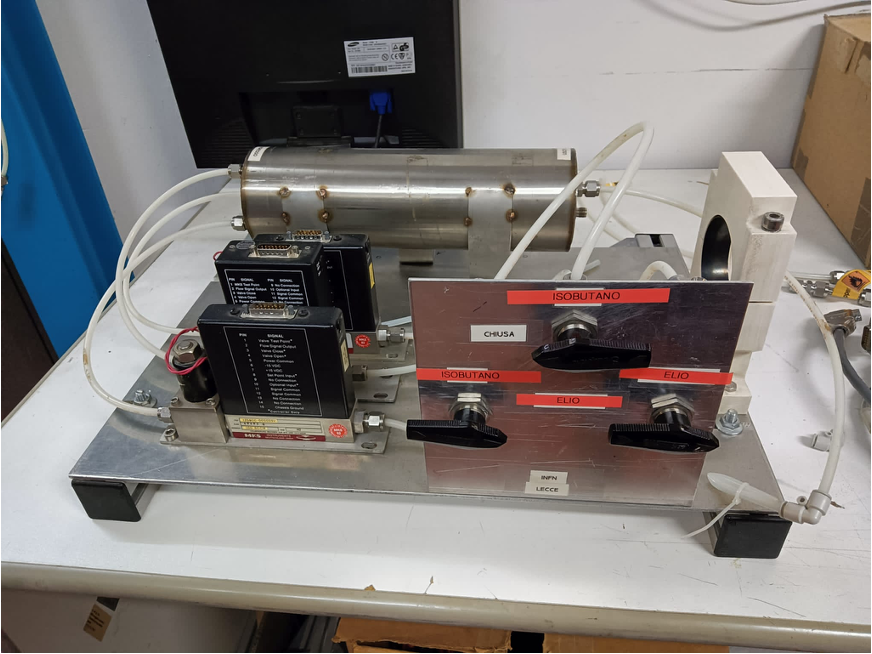}
\qquad 
\caption{Left: The set of drift tubes with different cell size used for the beam tests. Right: the portable gas system for gas flow monitoring and mixing.}
\label{Fig:Components}
\end{figure}

The experimental setup for a beam test performed in 2021, illustrated in Figure~\ref{Fig:Setup}, was designed as a fully integrated detection and readout system, a configuration that was subsequently adopted in later beam tests. A picture of the drift tubes used for the beam tests is in Figure~\ref{Fig:Components} (left). The drift tubes’ sense wires were connected to a high-voltage (HV) supply card at one end, while an HV decoupling termination card with a 330 $\Omega$ impedance, to match the drift tubes' characteristic impedance, was used at the other. Despite the relatively long coaxial cables (40 cm in 2021 and 60 cm in 2023), no preamplifiers were employed.

\begin{figure}[ht]
\centering

\includegraphics[width=.49\textwidth] {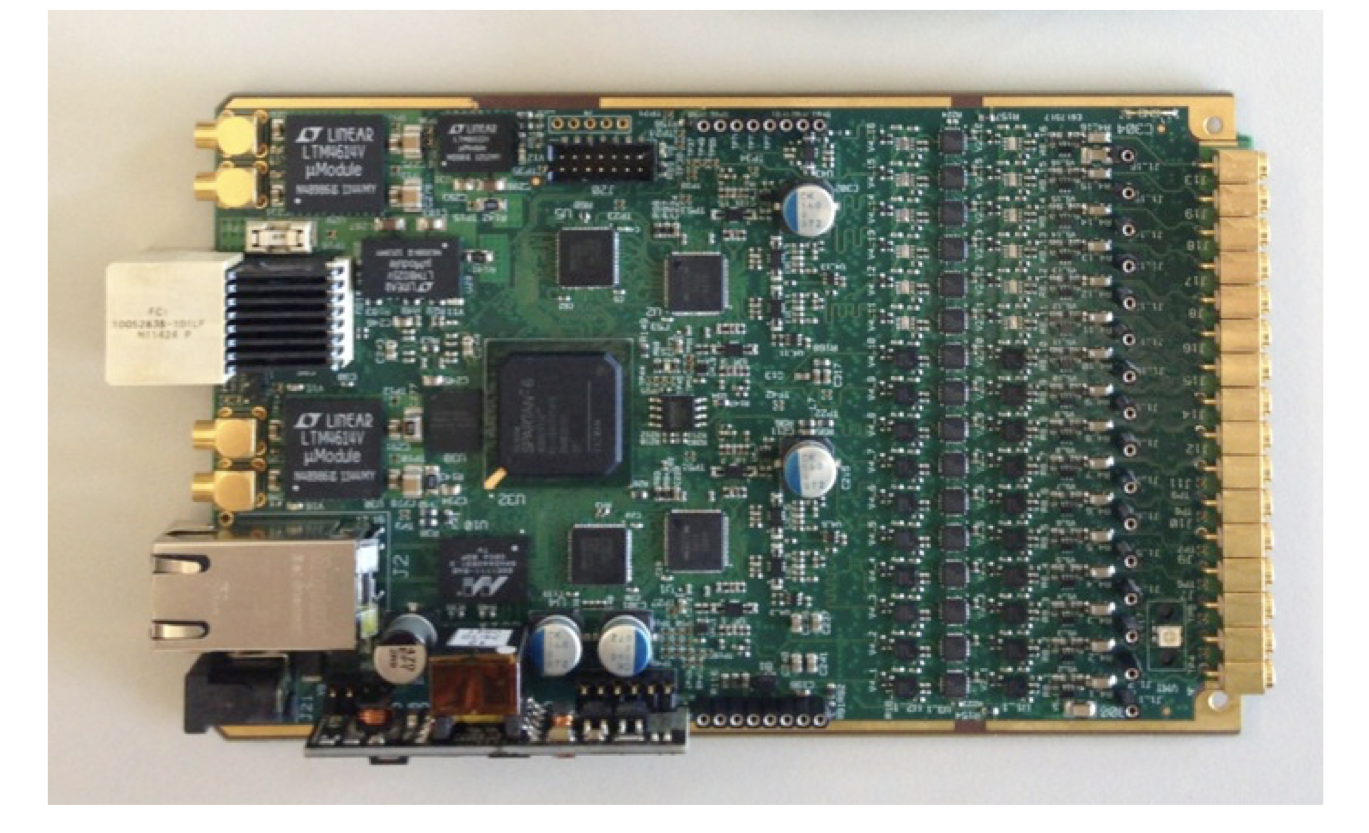}
\includegraphics[width=.49\textwidth] {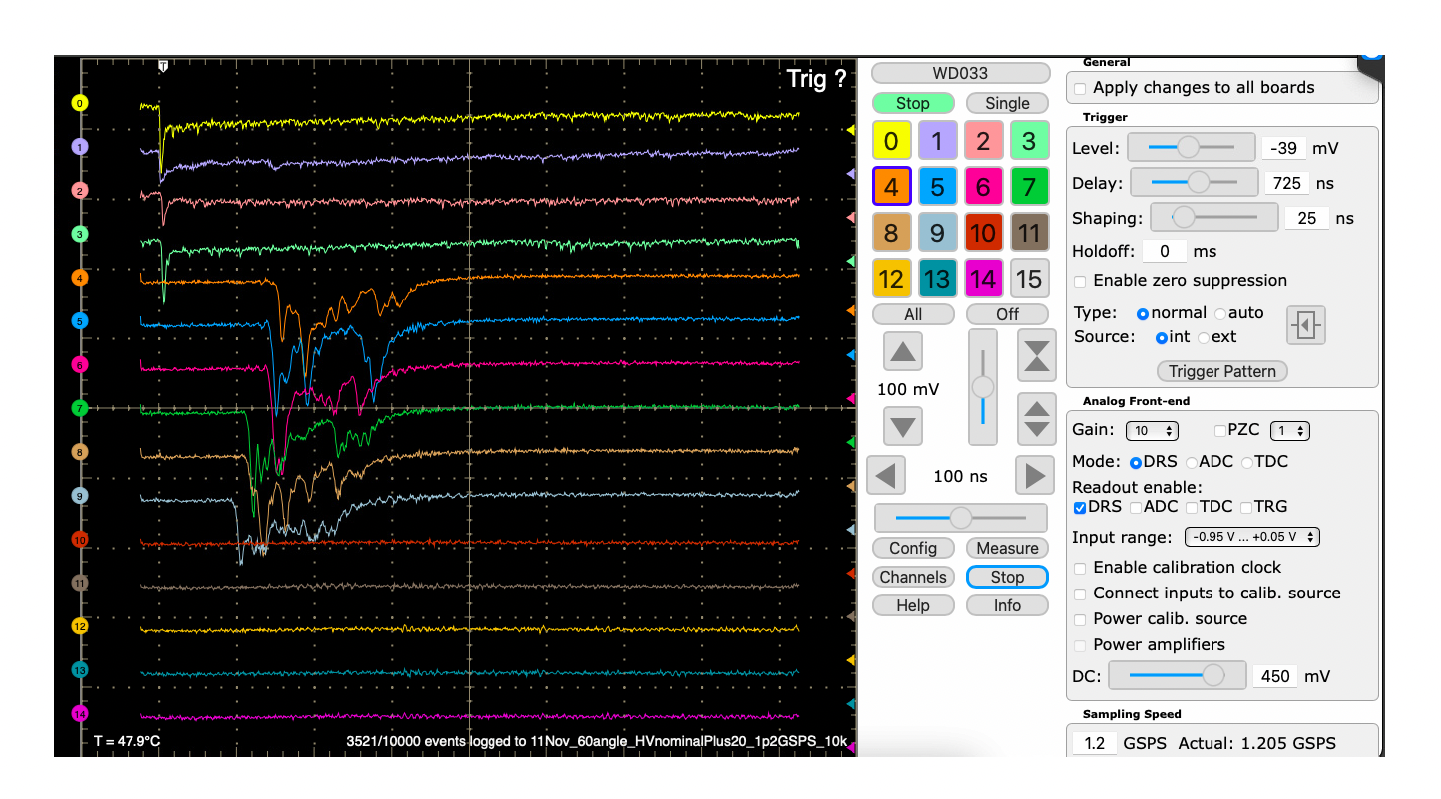}
\qquad 
\caption {The Wave Dream Board (WDB) (left) and its event display. On the right, the WDB control panel allows configuration of channel settings, including gain, threshold, bias voltage, and sampling rate, as well as access to the trigger definition panel.}
\label{Fig:WDB_1}
\end{figure}

The trigger system relied on a coincidence detection mechanism using scintillators coupled to silicon photomultipliers (SiPMs), ensuring precise event selection. A carefully controlled gas mixture, regulated by mass flowmeters under standard temperature and pressure (NTP) conditions, maintained stable detector operation; a picture of the portable gas system used for the gas flow monitoring and mixing is reported in Figure~\ref{Fig:Components} (right). The Wave Dream Board (WDB)\cite{Baldini_CL}, depicted in Figure~\ref{Fig:WDB_1} (left), served as the primary data acquisition unit. Featuring 16 channels with variable gain amplification and programmable pole-zero cancellation, the WDB provided flexible signal processing capabilities, though these features remained unused during beam tests. Two onboard DRS4 chips, each linked to an 8-channel ADC, enabled digitization at 80 MSPS with 12-bit resolution. Operating in “transparent mode,” the DRS4 continuously sampled input signals at up to 5 GSPS using an analog ring buffer, while an FPGA applied complex trigger algorithms based on threshold cuts on the summed input signals. The WDB functioned autonomously, with data read out via Gigabit Ethernet, and an ultra-low-noise bias voltage generator ensured stable operation of the SiPMs.

This experimental setup offered several key advantages. The availability of muon beams with varying $\beta\gamma$ values facilitated particle identification (PID) studies across a broad momentum range. The use of stacked metal drift tubes with a low material budget effectively mitigates multiple scattering effects, thereby allowing for precise ionization measurements. External tracking systems were unnecessary, as the drift path length within the tube was well-defined. Moreover, the design eliminated the need for internal tracking, thereby removing complexities associated with t$_{0}$ calibration, alignment, and time-to-distance conversion. The analysis relied solely on counting ionization clusters in the time domain, streamlining data interpretation and enhancing the overall efficiency of the measurement process.

Figure~\ref{Fig:WDB_1} (right) presents an event display from the WDB, which functions similarly to an oscilloscope. The first four channels correspond to the trigger scintillators, read out by SiPMs and internally processed by the WDB. The subsequent six channels capture signals from a row of 1 cm drift tubes impacted by the same muon.

Beyond assessing particle identification performance using both dE/dx and dN/dx, the beam tests focused on understanding the fundamental aspects of cluster counting and optimizing its methodology. A key objective was to confirm the Poisson nature of cluster counting and investigate its dependence on the $\beta\gamma$ of the ionizing track. Additionally, the tests aimed to determine the most effective cluster counting algorithms by exploring various parameters, including geometrical configurations, gas mixtures, gas gain levels, electric field strengths, and the performance of frontend and digitization electronics. Another critical aspect of the study was identifying the factors that could potentially limit cluster counting efficiency, such as cluster dimensions, space charge distribution, and electron-ion recombination effects.

To achieve these goals, multiple sets of square cross-section brass drift tubes were tested, with sizes ranging from 1.0 cm to 3.0 cm. These tubes were equipped with sense wires of varying diameters, from 10 to 40 $\mu$m, and operated at gas gains between 1 and 5 $\times 10^{5}$. Different helium-isobutane mixtures (90/10, 85/15, and 80/20) were employed to study the impact of gas composition on the detection process. The experimental setup was exposed to muon beams with different momenta across three beam test campaigns, conducted over the past three years.

\begin{figure}[ht]
\centering
\includegraphics[width=\textwidth] {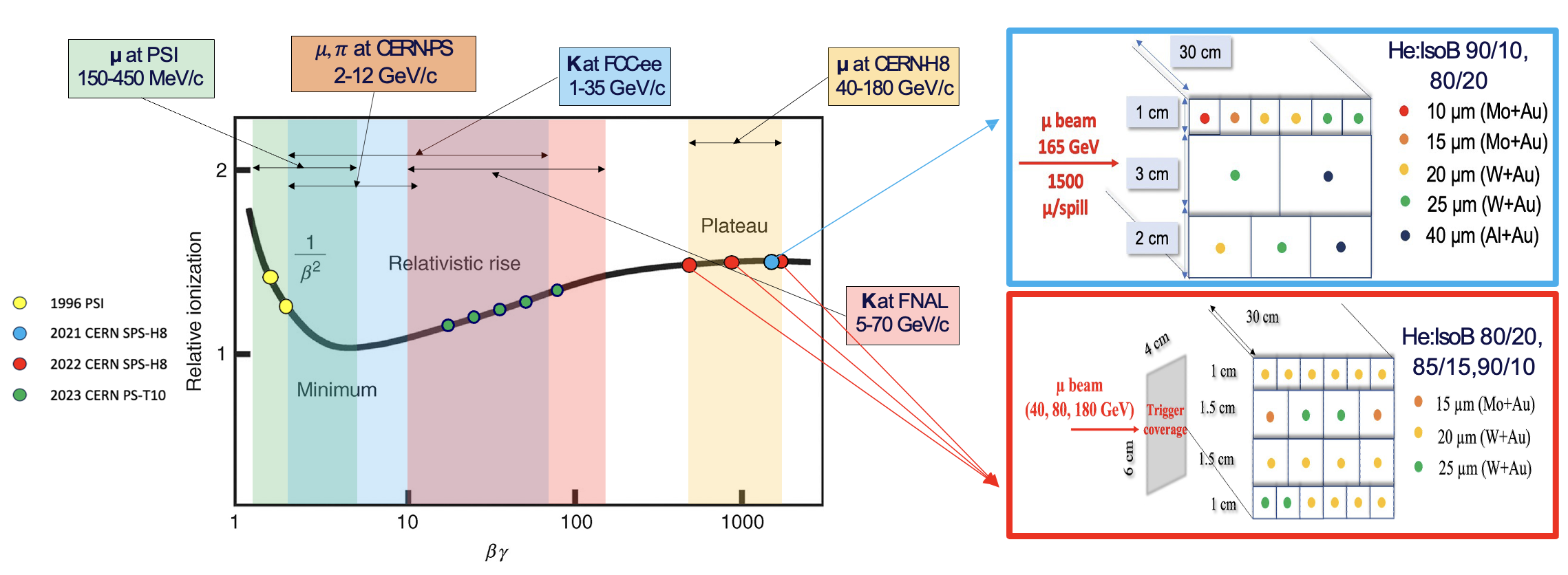}
\caption{Beam test campaigns with muons of the indicated $\beta\gamma$ values superimposed to a generic Bethe-Bloch curve. The explored $\beta\gamma$ values cover the regions of interest for particle identification at the future lepton colliders.}
\label{Fig:Rel}
\end{figure}

The initial tests took place at CERN SPS-H8 in November 2021 and July 2022, where muons with momenta between 40 and 180 GeV/c were used, corresponding to the Fermi plateau region. Subsequent tests were carried out at CERN PS-T10 in July 2023 and July 2024, focusing on the relativistic rise region, where muon momenta ranged from 2 to 10 GeV/c. These campaigns, summarized in Figure~\ref{Fig:Rel}, were complemented by pioneering measurements performed during an earlier beam test at PSI~\cite{CC}. Collectively, these tests provide crucial experimental validation of the cluster counting technique and contribute to refining particle identification methodologies for future applications.

\section{Results}

Establishing the operating parameters of the various drift tubes under different conditions (especially the gas gain, which influences the single-electron pulse height) is a critical preliminary step to ensure uniform data collection for assessing the efficiency of different cluster counting algorithms.

\subsection{Gas gain scan}

The gas gain as a function of anode high voltage for each drift tube configuration was determined by fitting the single-electron pulse height distribution with a Landau function. 
The extracted most probable value (MPV) was then corrected for the amplifier gain, mismatch between the characteristic tube impedance and the termination resistor, and the current divider effect between the termination resistor and the ADC input impedance.

\begin{figure}[ht]
\centering
\includegraphics[width=\textwidth] {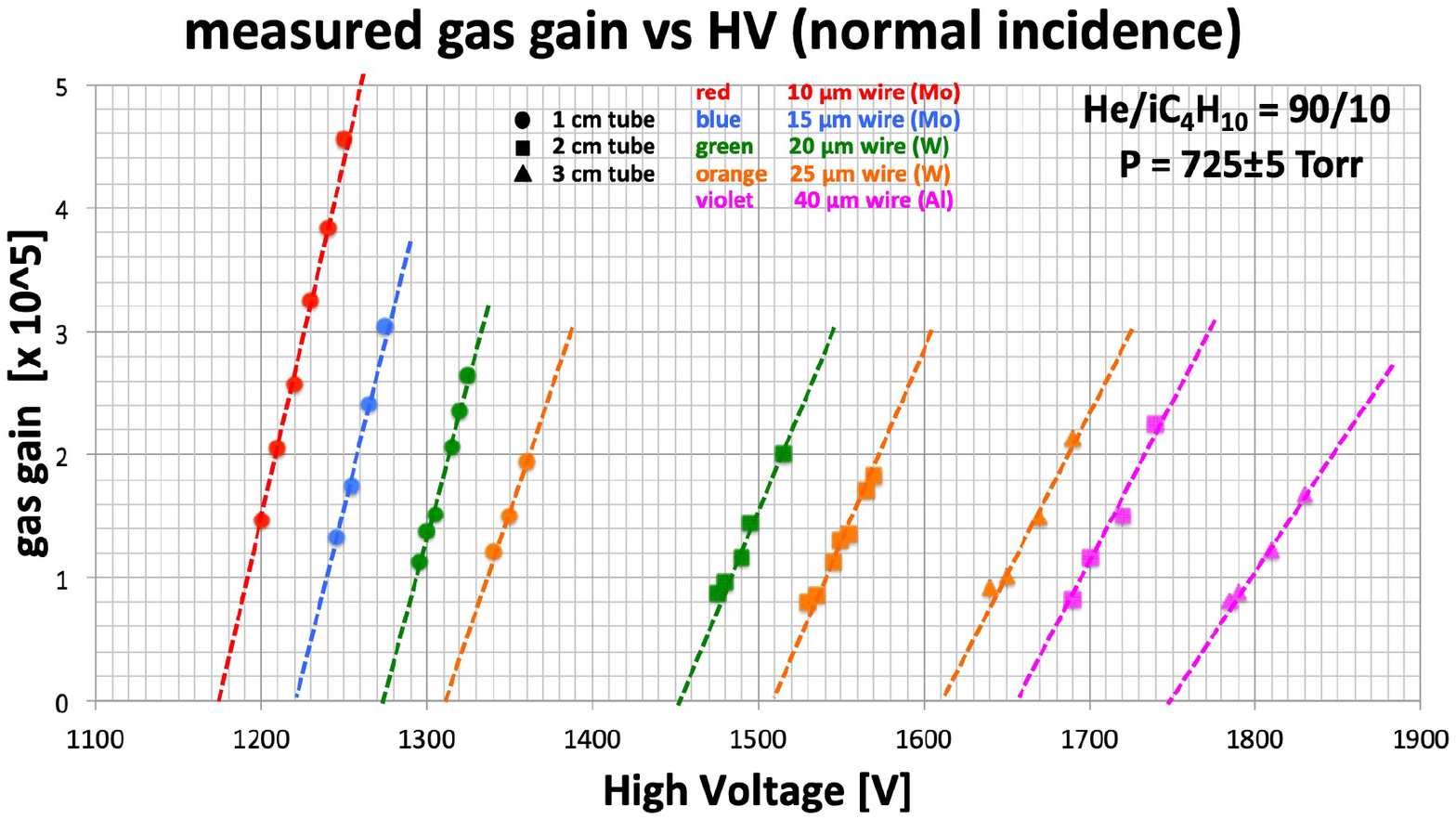}
\caption{Gas gain as a function of anode high voltage, derived from the November 2021 beam test data. The gas mixture and absolute operating pressure are specified. Different drift tube sizes are represented by distinct symbols, while various sense wire diameters are indicated by different colors. The dashed lines serve as a visual guide.}
\label{gain_Nov}
\end{figure}
\begin{figure}[ht]
\centering
\includegraphics[width=\textwidth,height=0.4\textheight] {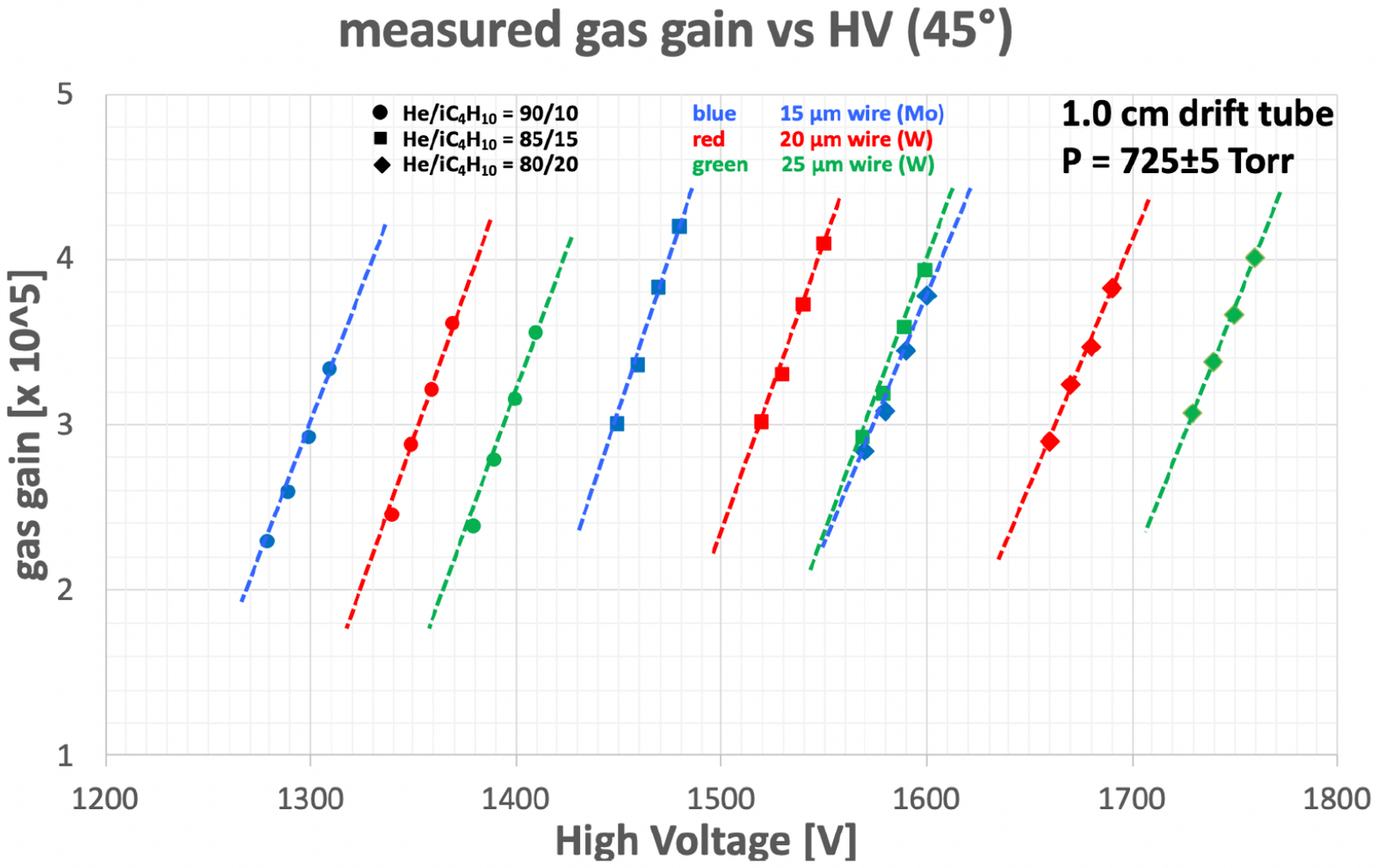}
\includegraphics[width=\textwidth,height=0.4\textheight] {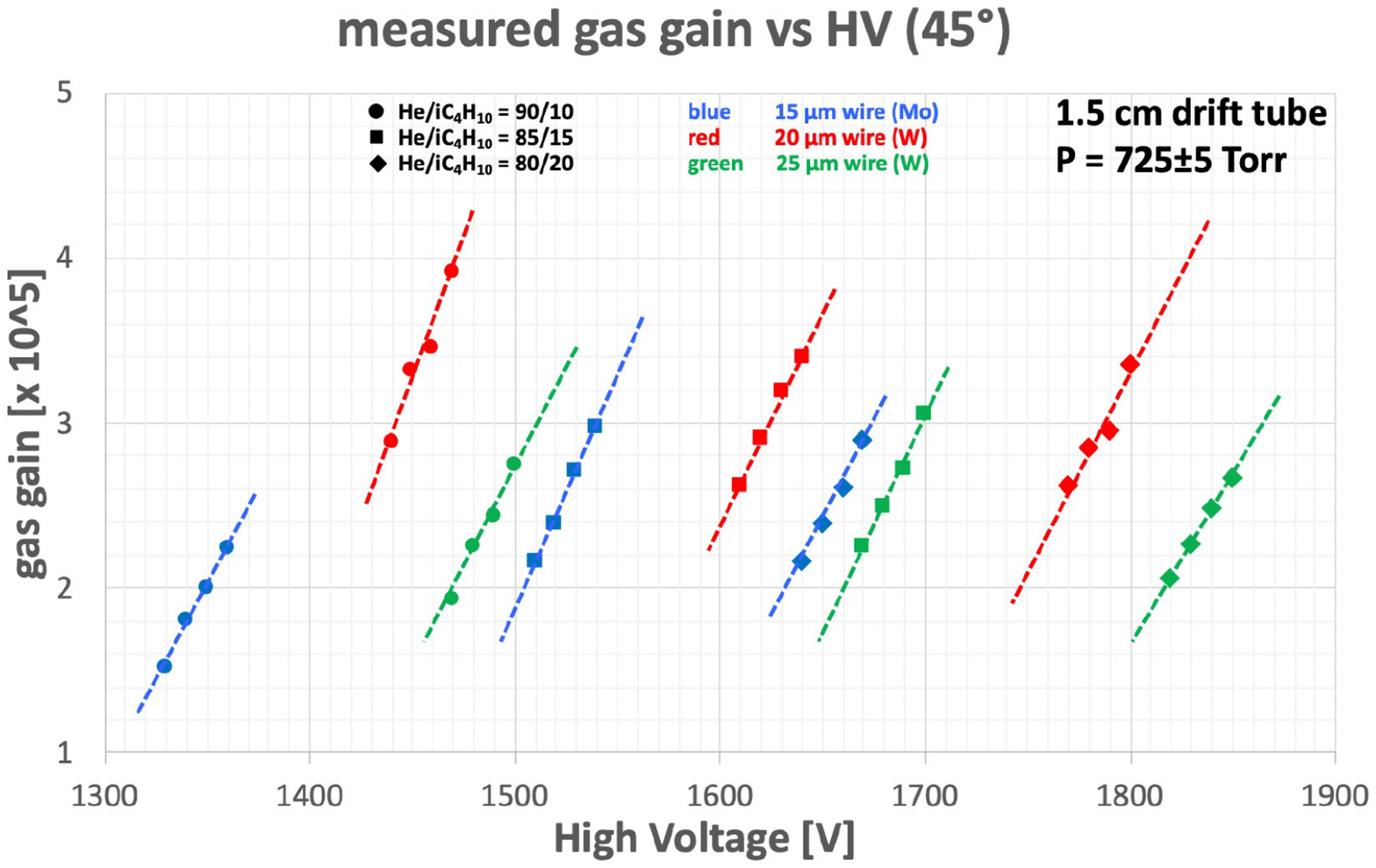}

\caption{
Gas gain as a function of anode high voltage, based on the July 2022 beam test data for drift tubes with 1 cm (top) and 1.5 cm (bottom) sizes. The 45° angle represents the angle between the beam direction and the normal to the sense wire. The absolute gas operating pressure is specified. Different gas mixtures are indicated by distinct symbols, while various sense wire diameters are shown in different colors. The dashed lines are provided for visual guidance.}
\label{gain_Jul}
\end{figure}

Figure~\ref{gain_Nov} presents the gas gain as a function of anode high voltage for all drift tube configurations tested during the 2021 beam test, using a 90\% He – 10\% iC$_{4}$H$_{10}$ gas mixture at an absolute pressure of 725 Torr, with data collected at normal beam incidence to the sense wires. Similarly, Figure~\ref{gain_Jul} illustrates the corresponding measurements from the 2022 beam test, which involved drift tubes of different sizes and various gas mixtures.

Optimal implementation of the cluster counting technique requires a gas gain—regardless of drift tube configuration (drift length, sense wire diameter, or gas mixture)—within the range of 1$\times10^{5}$ to 5$\times10^{5}$.

\subsection{Peak-Finding methods for electron detection}
\label{Sec:Algorithms}

The analysis presented in this paper employs two distinct peak-finding methods to identify electrons in drift signal waveforms. Prior to applying these methods, a preprocessing step is performed to establish a zero-baseline for each waveform. This is achieved by subtracting the mean pulse height calculated over the first 30 ns, i.e. an interval outside the signal acceptance window. The root mean square (r.m.s.) of the pulse height over this interval is then used to define the noise level, typically of the order of 1 mV.

The first method, the \textbf{Derivative-Based Algorithm (DERIV)}, identifies electron peaks by analyzing the first and second derivatives of the waveform. The first derivative is computed as the difference in amplitude between consecutive bins, while the second derivative is obtained as the difference of the first derivative over the same bins. A peak candidate is identified when the first derivative at its position falls below a predefined threshold, proportional to the r.m.s. noise, and exhibits a rising-falling pattern within a margin also scaled by the noise level. Additionally, the second derivative is used to verify waveform concavity at the peak position. Further selection criteria include amplitude thresholds, based on the r.m.s., and constraints on the amplitude difference between the peak candidate and its neighboring bins.

\begin{figure}[ht]
\centering
\includegraphics[width=\textwidth]{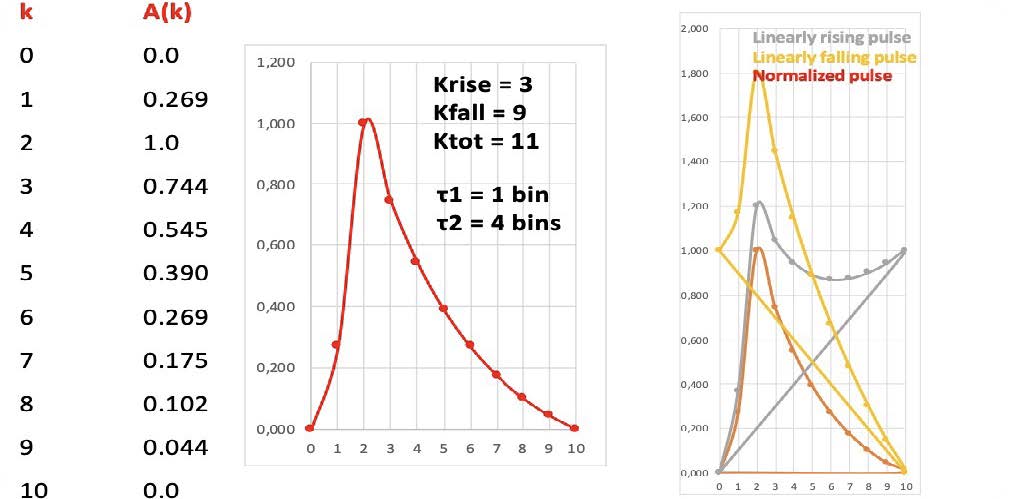}
\qquad 
\caption{Left: Example of an electron pulse template used by the RTA algorithm to scan and identify waveform peaks. Right: Normalized pulse template overlaid on a rising (grey) and a falling (yellow) edge of the waveform.}
\label{RTA_algo}
\end{figure}

The second method, the \textbf{Running Template Algorithm (RTA)}, employs a predefined electron pulse template, characterized by rising and falling exponentials, which is derived from experimental data and digitized according to the sampling rate, as illustrated in Figure~\ref{RTA_algo}. The algorithm scans the waveform within a search window, comparing the normalized template to the data. A match is identified when the agreement exceeds a predefined threshold. Once a peak is detected, it is subtracted from the waveform and its position and amplitude are stored. The search continues to the rest of the waveform and the process is repeated iteratively until no further peaks are found.

\begin{figure}[ht]
\centering
\includegraphics[width=14cm] {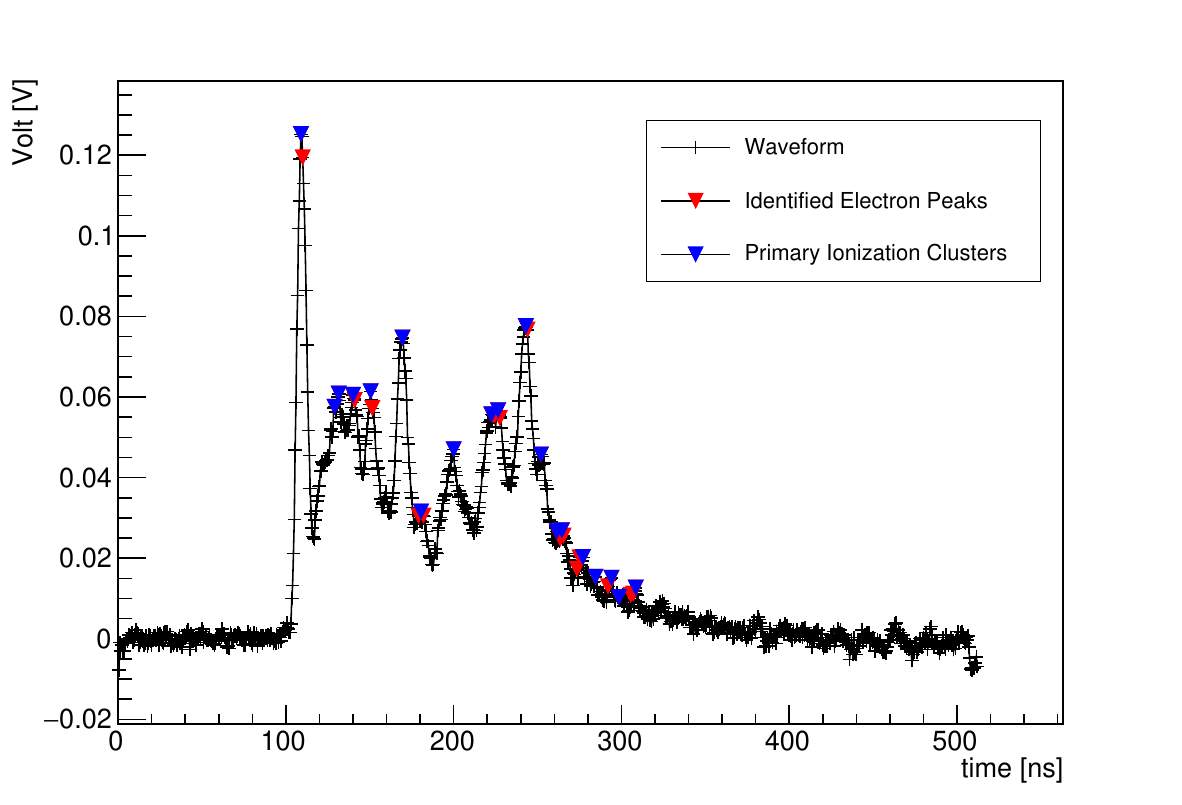}
\includegraphics[width=14cm] {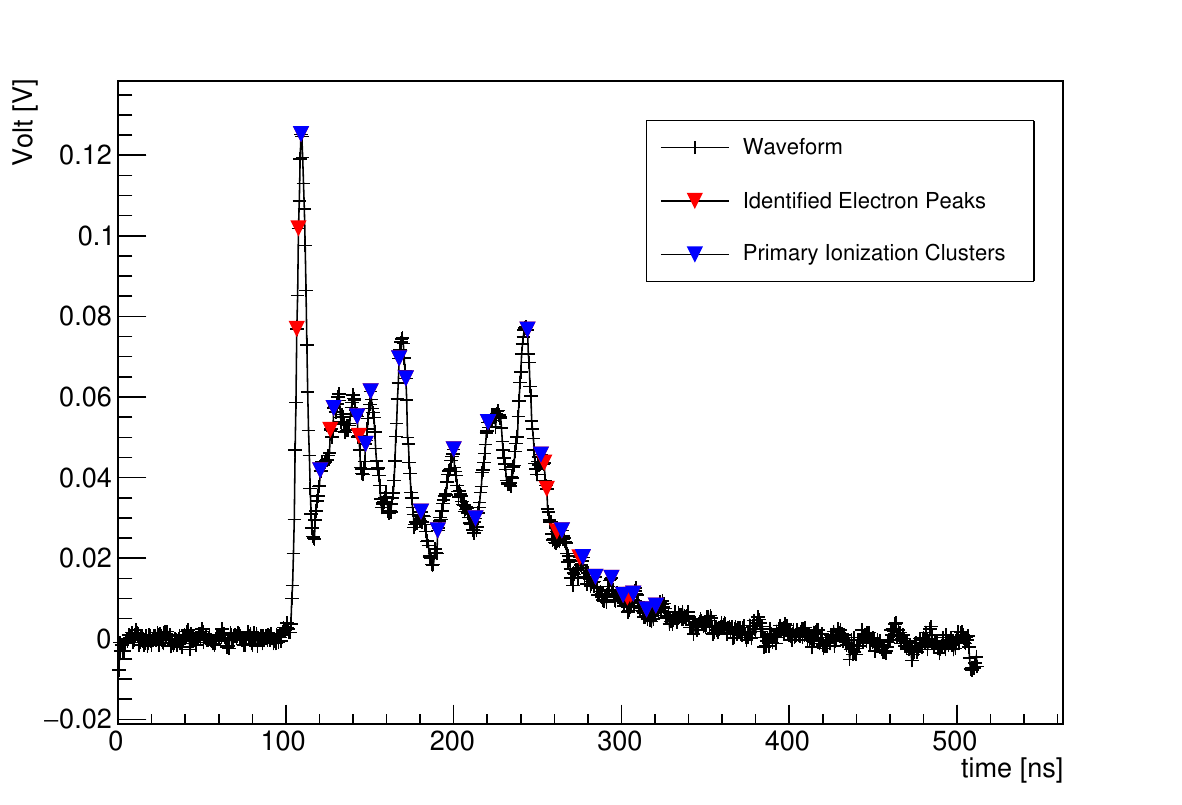}
\caption{Recorded waveform for a 1 cm drift cell with a 20 $\mu$m sense wire diameter, a $45^{\circ}$ track angle, a sampling rate of 2 GSa/s, and a He/iC$_{4}$H$_{10}$ 90/10 gas mixture. Electron peak identification is performed using the RTA algorithm (top) and the DERIV algorithm (bottom), where blue arrows indicate cluster peaks and red arrows represent individual electron peaks.}
\label{Fig:Waveform}
\end{figure}

Figure~\ref{Fig:Waveform} illustrates, as an example, the peaks identified in a drift signal waveform by both algorithms. The results indicate that, with proper tuning, both approaches can deliver comparable accuracy in identifying ionization clusters. However, the RTA algorithm is adopted in the following analyses, as it demonstrates more reliable peak detection performance, particularly in complex or noisy waveforms, even though the mean values obtained from the two algorithms are comparable. Additionally, machine learning (ML) techniques have shown great potential in enhancing ionization cluster identification and will be explored further in future studies~\cite{ML}.

\subsection{Identification of Primary Ionization Clusters}
\label{Sec:Cls}

The process of identifying primary ionization clusters, known as clusterization, involves grouping together all ionization electrons originating from the same ionization event. This procedure associates the peaks detected in the signal waveform by the previously described algorithms into clusters, assigning a drift time and amplitude to each.

The clustering process follows a systematic approach: electron peaks that are temporally contiguous and exhibit time differences compatible with diffusion-induced spreading are grouped into the same ionization cluster. The total number of electrons within each cluster is then determined accordingly. The cluster’s drift time and amplitude are assigned based on the electron peak with the highest amplitude within the group. An example of identified clusters, highlighted by blue markers, is shown in Figure~\ref{Fig:Waveform}. Slight differences in the results of the two algorithms reflect the peculiarity of the two different methods.

The number of identified clusters obtained using the RTA combined with the clusterization algorithm is consistent with the expected values with a slight decrease due to the recombination and attachment effect that will be explained latter in section sec.~\ref{corrClusters}. In a 1.0 cm drift tube with a drift length of 0.8 cm (effectively extending to 1.13 cm due to the 45° beam angle relative to the normal to the sense wire), the expected number of clusters is 18.4. This estimate is obtained assuming a cluster density of 12.5 clusters per cm in a 90\% He – 10\% iC$_{4}$H$_{10}$ gas mixture for a minimum ionizing particle (m.i.p.)~\cite{CC}, a Fermi plateau correction factor of 1.3 relative to a m.i.p.

\begin{figure}[ht]
\centering
\includegraphics[width=\textwidth] {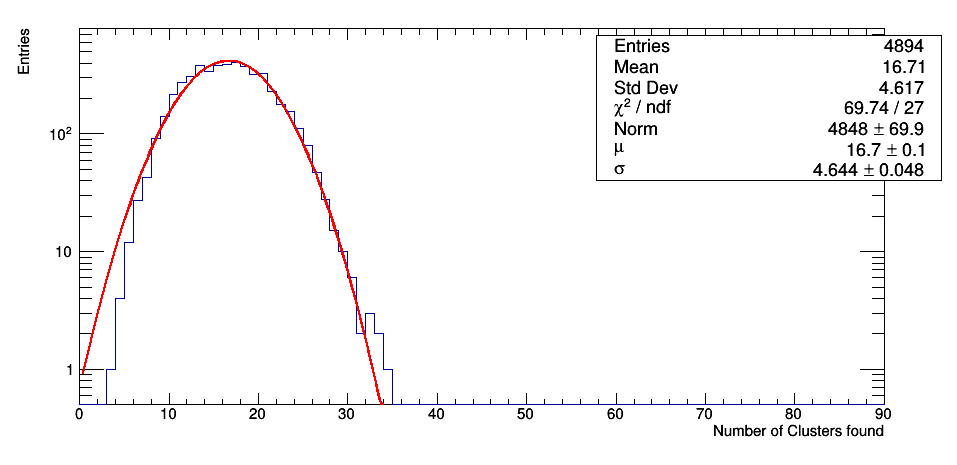}
\includegraphics[width=\textwidth] {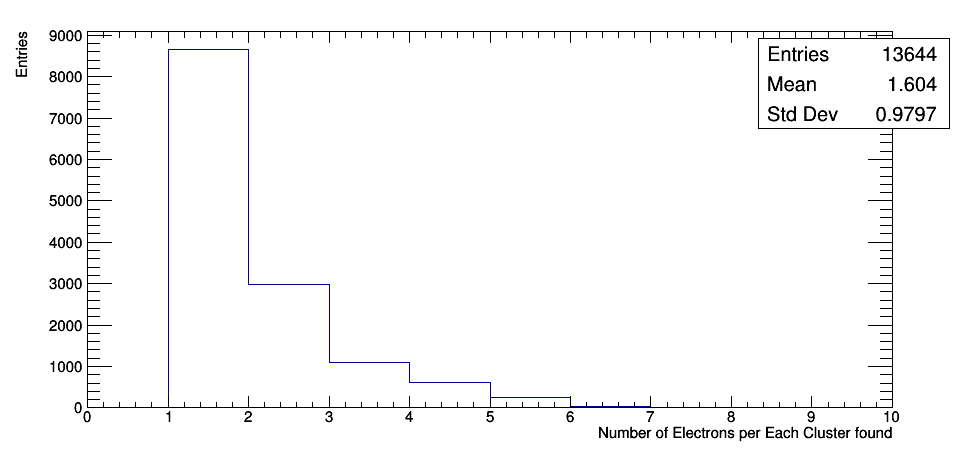}
\caption{Distribution of the number of clusters (top) and the number of electrons per cluster (bottom) for a 1 cm drift cell with a 20 $\mu$m sense wire diameter, a $45^{\circ}$ track angle, a sampling rate of 2 GSa/s, and a He/iC$_{4}$H$_{10}$ (90/10) gas mixture, obtained using the RTA combined with the clusterization algorithm.}
\label{Fig:Clusters}
\end{figure}

Figure~\ref{Fig:Clusters} (top) presents the cluster distribution obtained using the clustering algorithm. A Poisson fit, represented by the red curve, shows a mean value consistent with theoretical expectations. To validate the reliability of the applied methodology, consistency checks are performed by analyzing the electron population per cluster. Figure~\ref{Fig:Clusters} (bottom) illustrates the average number of electrons per cluster, demonstrating good agreement with experimental measurements~\cite{CLS}. Large cluster sizes are not present in the plot because of the low probability of appearance and the very limited statistics examined.

\subsection{Cluster Counting Performance under Different Measurement Conditions}
Using test beam data, we evaluated the performance of our algorithms under various operational conditions, including different gas mixtures, gas gain values, geometrical configurations (such as cell size and sense wire dimensions), sampling rates, high voltage (HV), and track angles. Additionally, we investigated the impact of selecting templates with different rising and falling exponentials for the RTA algorithm.

This section presents a comparative analysis of different RTA templates and assesses the optimized template against the DERIV algorithm. Furthermore, it provides a detailed evaluation of cluster counting algorithm performance, highlighting the influence of key parameters such as gas gain, ionization path length, and gas composition on the drift chamber’s ability to detect ionization clusters. Nominal HV value corresponds to 2$\times10^{5}$ gas gain and a $\pm$10V difference from nominal HV corresponds approximately to a $\pm$20\% variation in gas gain.

\begin{figure}[ht]
\centering
\includegraphics[width=7.5cm]  {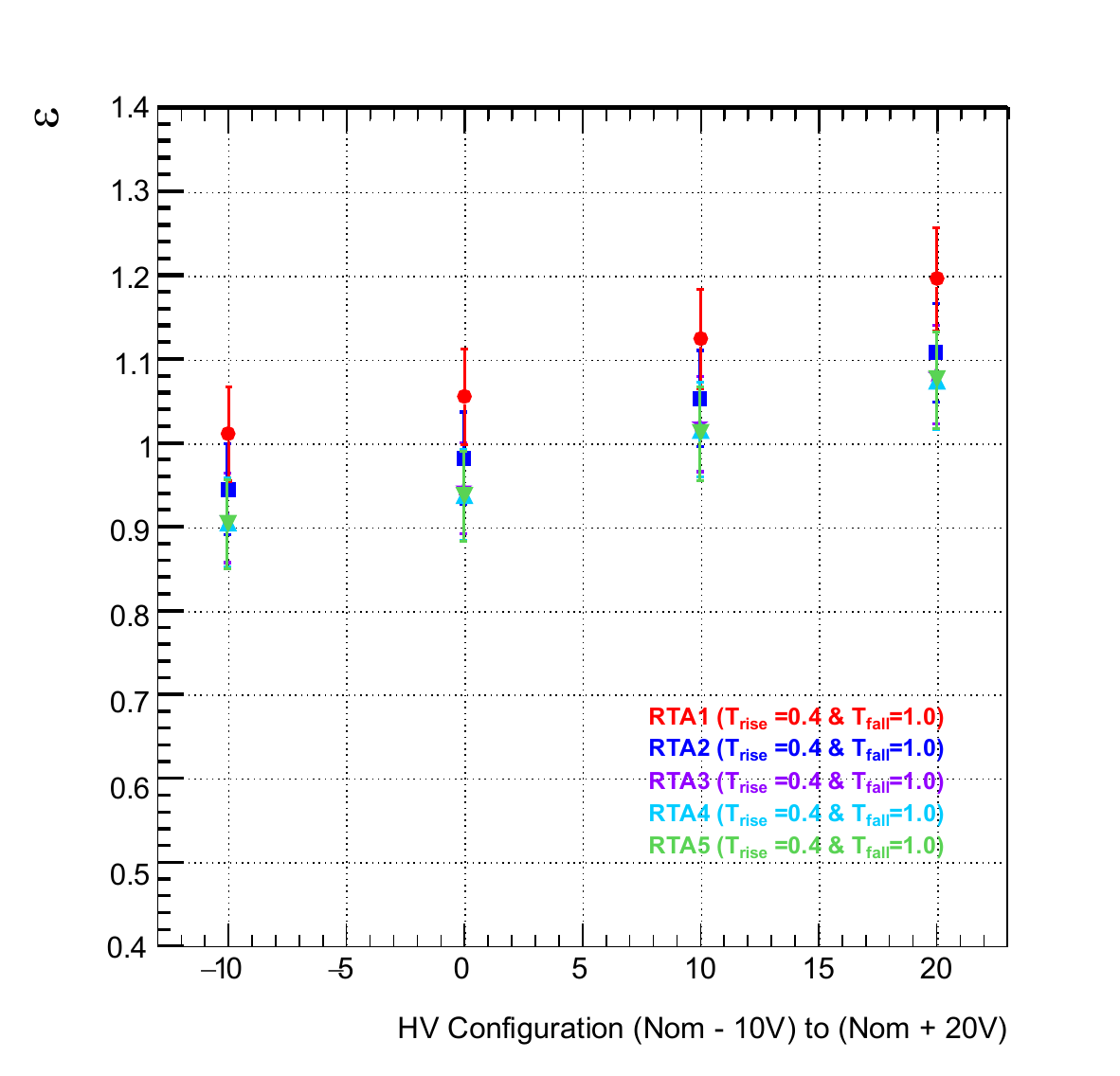}
\includegraphics[width=7.5cm]  {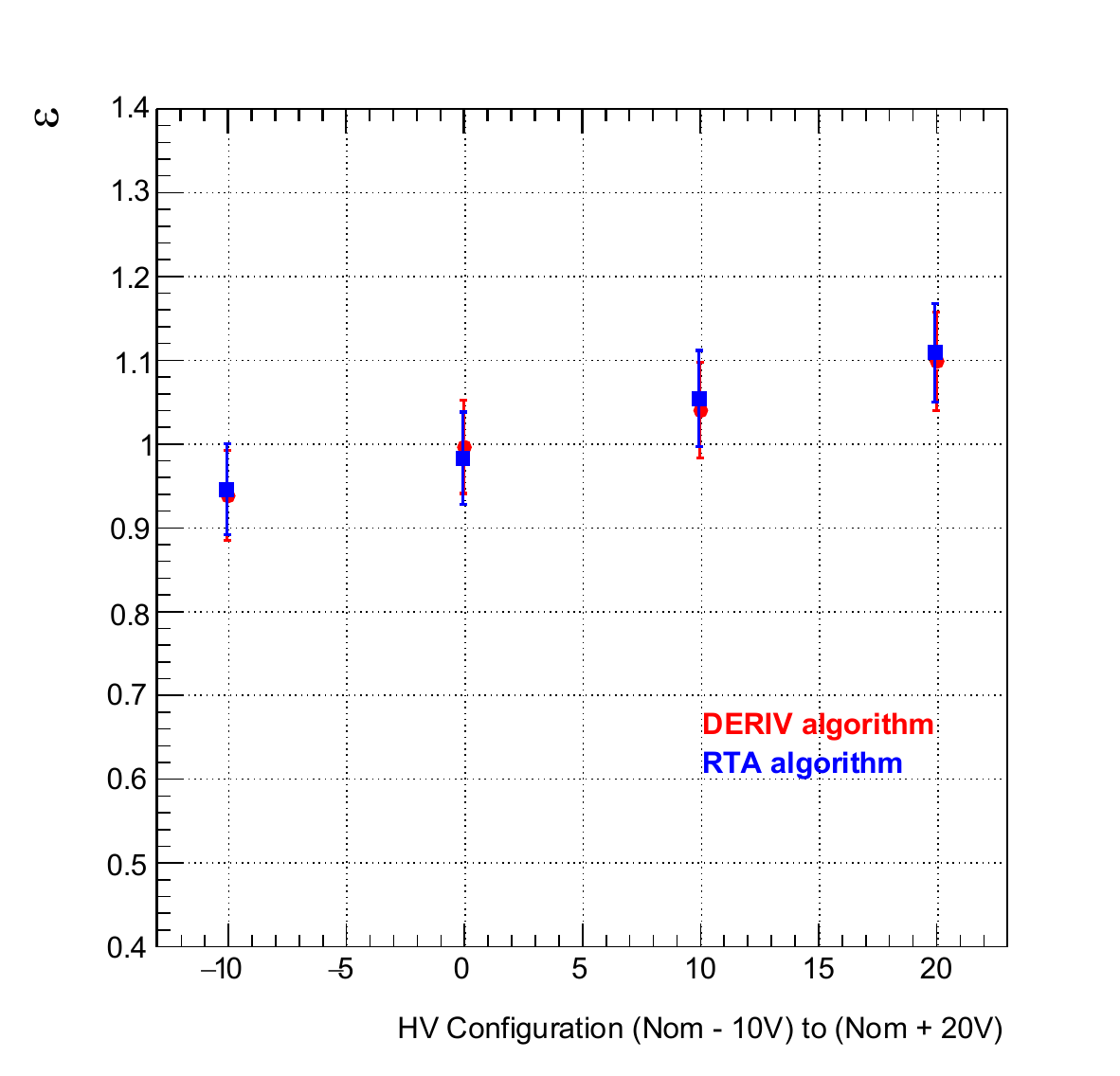}

\caption{Comparison of performance scans for different templates used in the RTA algorithm (left) and a comparison of the selected template's performance with that of the DERIV algorithm (right) as a function of drift tube high voltage configuration sets. The
uncertainties shown are statistical.}
\label{Fig:RTAScan}
\end{figure}

\subsubsection{RTA templates scan}
Figure~\ref{Fig:RTAScan} (left) illustrates a performance scan of the RTA algorithm using different electron pulse templates, characterized by varying rise (\(T_{\text{rise}}\)) and fall (\(T_{\text{fall}}\)) times. The measured average number of clusters, normalized to the expected number ($\varepsilon$), is shown as a function of the drift tube high voltage (HV). It is worth noticing here that the parameters of the RTA algorithm used in this analysis are optimized for nominal HV and, therefore, different HV, i.e., different gas gains, will affect the cluster finding performance, with effects confined to maximum variations of the order of 10\% in this range of gas gains. This is a further confirmation of the need of optimizing the cluster finding parameters as a function of the gas gain. 
The overall performance of all tested configurations remains consistent within the statistical uncertainties. This indicates a high degree of robustness in the RTA algorithm with respect to the choice of pulse template. Among the tested templates, the configuration with \(T_{\text{rise}} = 0.5\) and \(T_{\text{fall}} = 1.3\) (blue squares) provides the closest match to the expected number of clusters across all HV settings, reinforcing its suitability as a nominal choice.

Figure~\ref{Fig:RTAScan} (right) compares the performance of the selected RTA template (\(T_{\text{rise}} = 0.5\), \(T_{\text{fall}} = 1.3\)) with that of the DERIV algorithm. The two methods yield consistent results across the full range of HV configurations, with both algorithms maintaining an average cluster count consistent with the expected value. This agreement validates the reliability of the RTA approach while also demonstrating that, when appropriately tuned, both algorithms achieve comparable performance in cluster identification. 

\subsubsection{Cluster Counting Versus Ionization Density}
\begin{figure}[ht]
\centering
\includegraphics[width=7.5cm] {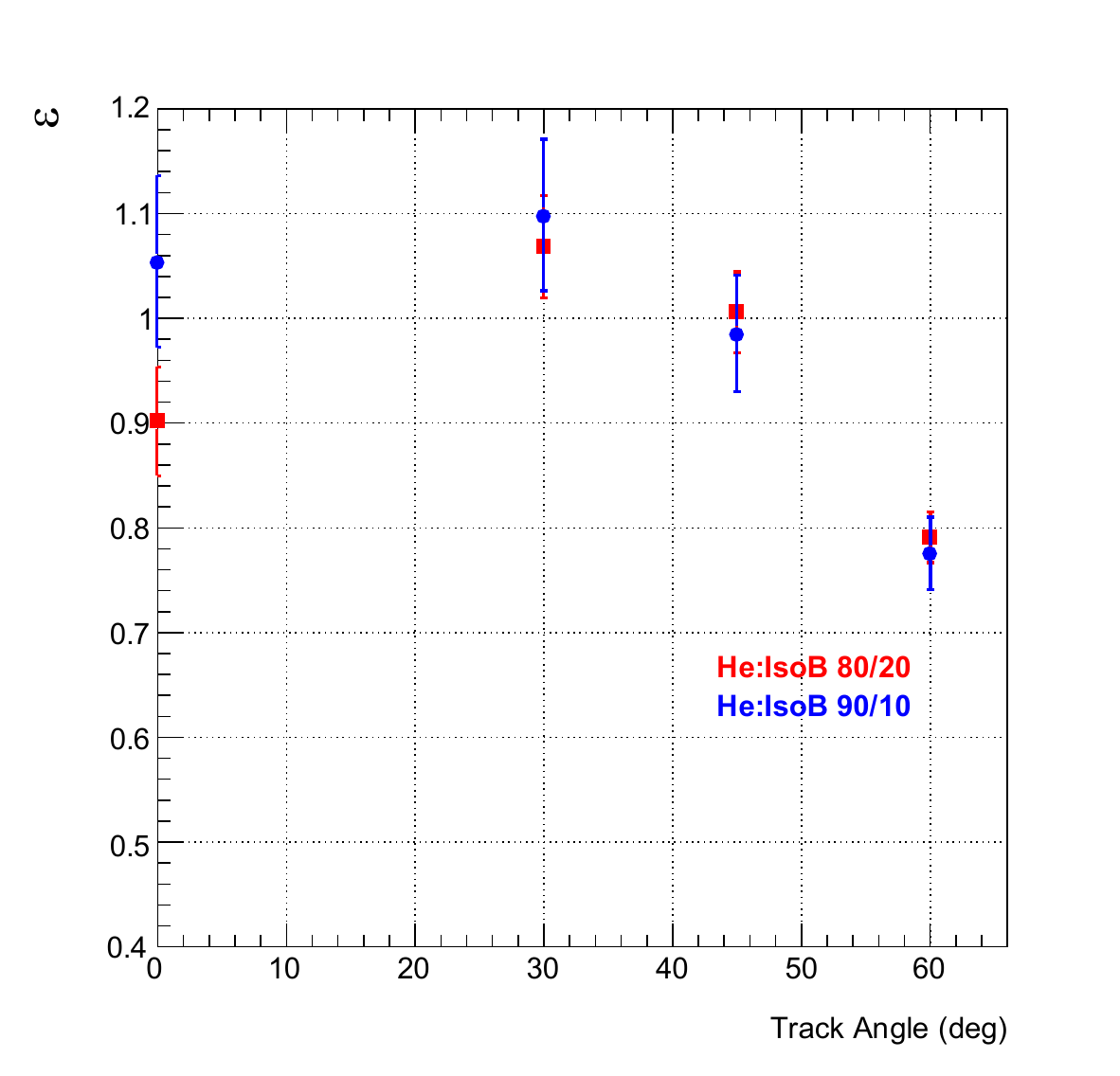}
\includegraphics[width=7.5cm] {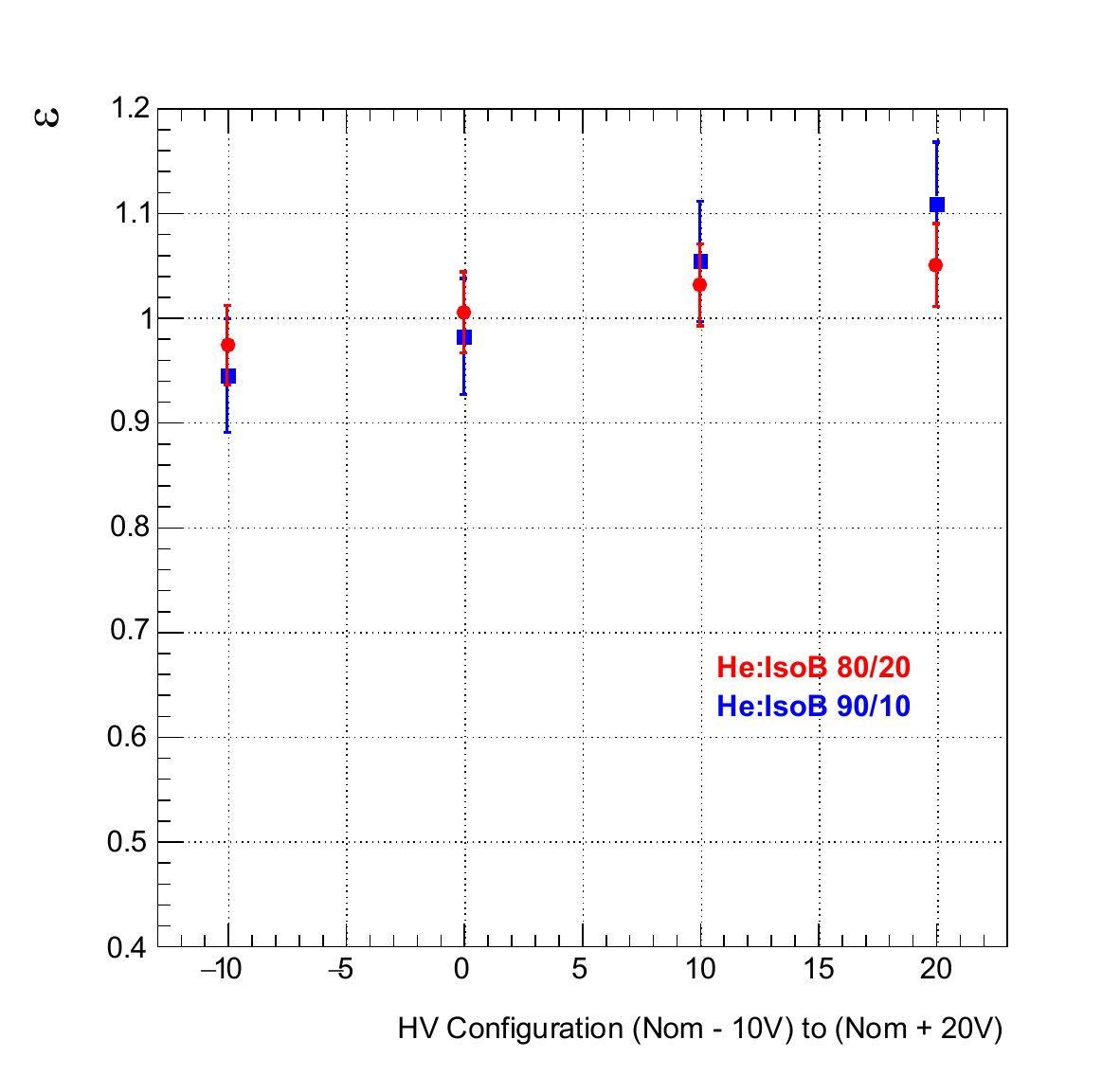}
\caption{Comparison of cluster counting performance as a function of track angle (left) and drift tube high voltage (right), obtained with the RTA + clusterization algorithm. Results are shown for two gas mixtures: 80/20 (red) and 90/10 (blue). The
uncertainties shown are statistical.}
\label{Fig:Scan}
\end{figure}

To account for variations in drift velocity across different gas mixtures, the clusterization cuts are optimized separately for each mixture. This ensures that the cluster counting efficiency remains consistent and accurate, irrespective of the specific gas composition. Figure~\ref{Fig:Scan} shows the cluster counting performance for two helium–isobutane mixtures: 90/10 and 80/20. The measured cluster densities—approximately 12.5 and 18 clusters per centimeter per minimum ionizing particle (m.i.p.), respectively—are in good agreement with the expected specific ionization values derived from empirical predictions. These results demonstrate that the ionization density of each gas mixture directly influences the cluster counting efficiency, confirming that variations in gas composition must be carefully considered when optimizing detection parameters.

\subsubsection{Cluster Counting Versus Ionization Length}

 In addition to ionization density, we explored the scaling behavior of cluster counts with track length by analyzing results at different angles between the beam direction and the normal to the wire direction. Figure~\ref{Fig:Scan} (left) presents the ratio of counted clusters to expected clusters as a function of beam angle for two different gas mixtures. A notable deficit in cluster counting is observed at normal incidence for the 80/20 helium-isobutane gas mixture. This is likely due to the onset of space charge effects, which distort the drift of ionization electrons and lead to inaccuracies in cluster counting particularly for normal incidence and for higher cluster densities. At larger beam angles, the performance of the cluster counting algorithms, optimized for 45° beam incidence, is affected by the increased cluster densities, requiring slight modifications in the algorithm parameters.

\subsubsection{Cluster Counting Versus Gas Gain}
Figure~\ref{Fig:Scan} (right) illustrates the relationship between cluster counting performance and gas gain. As in Figure~\ref{Fig:RTAScan} (right), each step in high voltage (10V) corresponds to a change of approximately 20\% in the gas gain, highlighting the sensitivity of the detection system to variations in gain. A slight dependence of counting efficiency on gas gain is observed, demonstrating that fluctuations in gas gain can influence the accuracy of cluster counting. This underscores the need for precise calibration and optimization of the detector parameters. All peak-finding parameters in this analysis were optimized for the nominal HV value corresponding to  2$\times10^{5}$ gas gain. Any observed undercounting or overcounting in the cluster counts can be corrected by fine-tuning these parameters, ensuring that the overall performance remains stable even in the presence of gain fluctuations.

\subsection{Comparative Analysis of dE/dx and dN/dx Resolution}

This section presents a comparative analysis of the resolution achieved through two distinct methods for measuring particle interactions: the energy loss per unit length (dE/dx) and the ionization cluster density (dN/dx). Both methods utilize identical particle tracks composed of the same hits, ensuring a direct and fair comparison. The tracks are approximately reconstructed from individual hits along the drift cells. For example, a track of about 2 m corresponds to the sum of hits across 178 cells (each 1 cm wide) with a drift distance of 0.8 cm and a beam incidence angle of $45^{\circ}$, resulting in an effective track length of approximately 2 m.

\begin{figure}[ht]
\centering
\includegraphics[width=7.5cm] {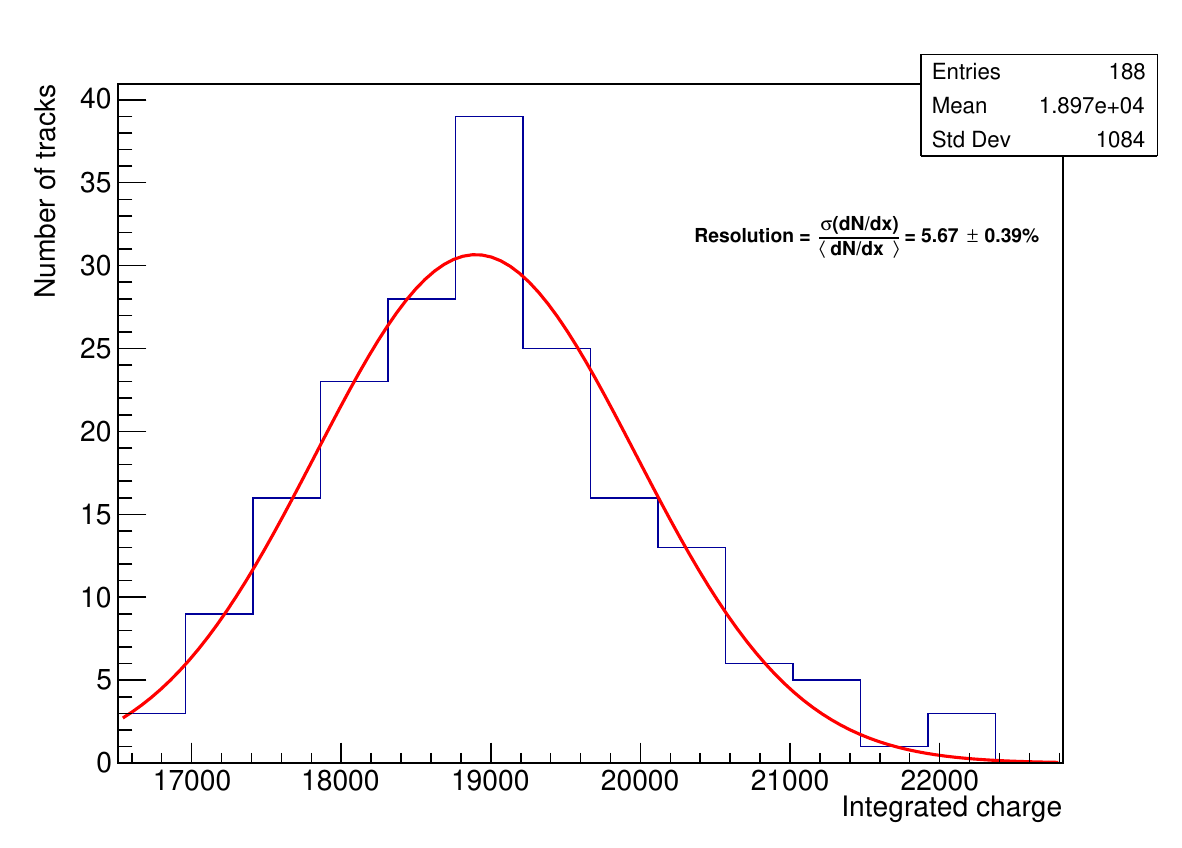}
\includegraphics[width=7.5cm]{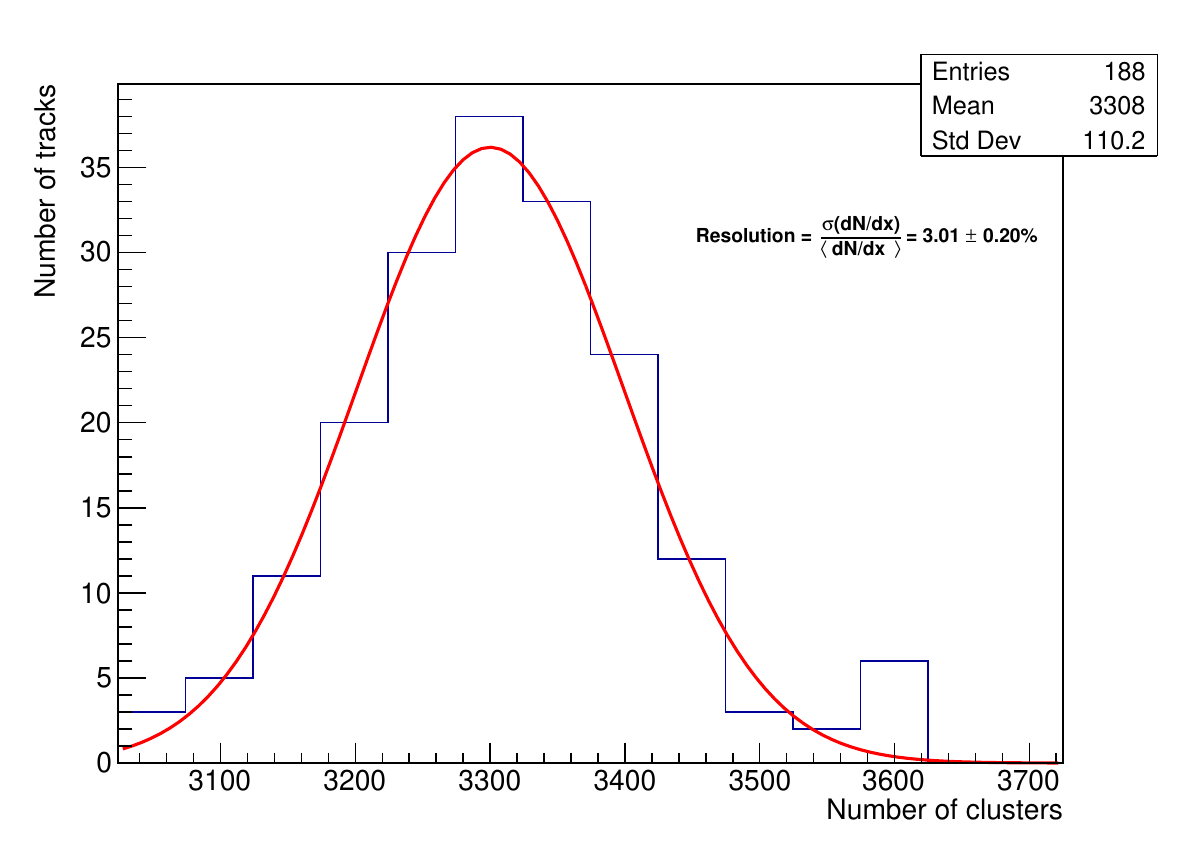}
\caption{Integral of the truncated mean charge distribution at 80\% (left) and the number of ionization clusters (right) measured along 2-meter tracks constructed from the same hits. The red curves represent the Gaussian fits to the data, with the quoted resolutions including the propagated fit uncertainties.}
\label{Fig:res_2m}
\end{figure}

\begin{figure}[ht]
\centering
\includegraphics[width=13.5cm]{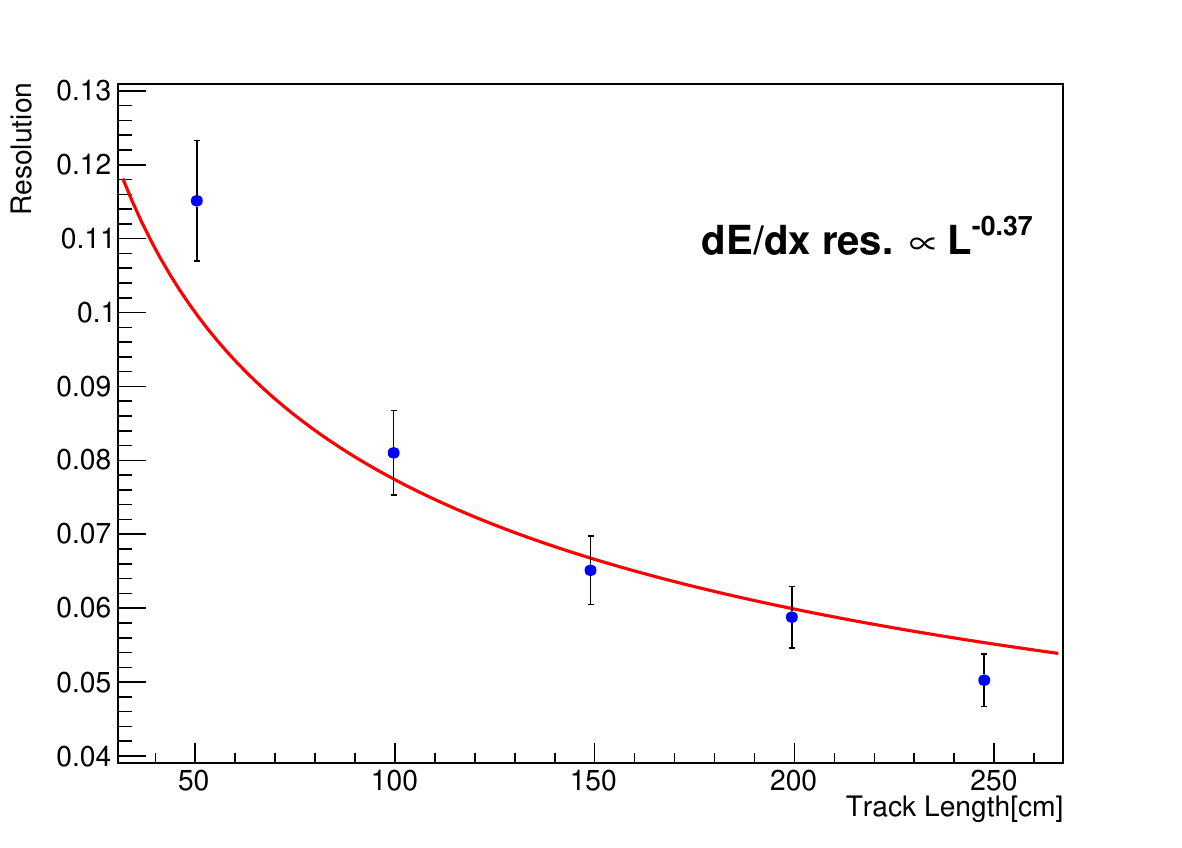}

\includegraphics[width=13.5cm]{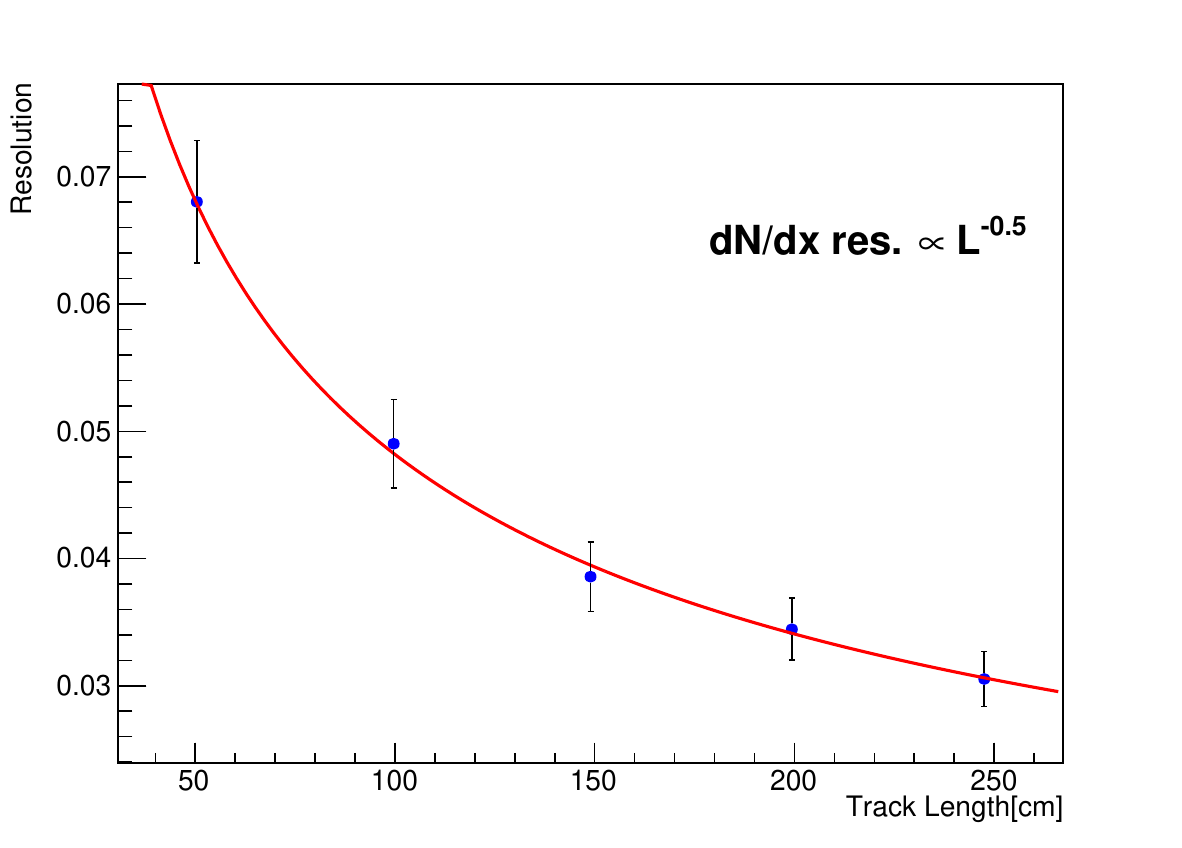}
\caption{Resolution as a function of track length (L) for the dE/dx method (top) and the dN/dx method (bottom). The red lines represent the fits with power-law dependencies L$^{-0.37}$ for dE/dx and L$^{-0.5}$ for dN/dx. The corresponding fit qualities are $\chi^{2}/\mathrm{NDF} = 1.56$ for dE/dx and $\chi^{2}/\mathrm{NDF} = 0.04$ for dN/dx.}
\label{Fig:res_scan}
\end{figure}

The energy loss along a particle's track, measured by dE/dx, follows a Landau distribution, characterized by a long tail due to high-energy $\delta$ electrons. To calculate the energy loss, the charge deposited in multiple drift cells along the track is recorded, and the mean charge of these samples is used as the dE/dx value. However, the simple mean is sensitive to large fluctuations arising from the nature of the Landau distribution, where outliers with high charge deposits can skew the result. To mitigate this, the truncated mean method is employed, discarding the highest charge samples—typically the top 20\% for He-based gas mixtures~\cite{Andryakov}—and calculating the mean of the remaining samples.

The resolution of dN/dx is assessed using the same tracks employed for the dE/dx analysis. The RTA + clusterization algorithm, described in Section~\ref{Sec:Cls}, is applied to reconstruct the ionization clusters. The truncated charge distribution is then compared with the number of ionization clusters measured along a 2-meter track length, as shown in Figure~\ref{Fig:res_2m}. The results demonstrate that the dN/dx method achieves a resolution of 3.01 $\pm$ 0.20\%, nearly twice as good as the 5.7 $\pm$ 0.39\% resolution obtained from the dE/dx method, in good agreement with both analytical calculations and simulation predictions. Furthermore, the resolution is evaluated for different track lengths, as shown in Figure~\ref{Fig:res_scan}, for both dE/dx and dN/dx methods. The resolution dependence on track length follows an L$^{-0.37}$ scaling for dE/dx, consistent with the Lehraus plot~\cite{Lehraus}, and an L$^{-0.5}$ scaling for dN/dx, highlighting the superior resolution provided by the cluster counting technique. The corresponding fit qualities are $\chi^{2}/\mathrm{NDF} = 1.56$ for dE/dx and $\chi^{2}/\mathrm{NDF} = 0.04$ for dN/dx.

\subsection{Improved dN/dx Resolution}
To improve the resolution achieved with the cluster counting technique (dN/dx), we applied two key procedures: waveform cleaning to reject distorted signals, and a time-based correction to account for recombination and attachment effects during electron drift.

\begin{figure}[ht]
\centering
\includegraphics[width=7.5cm] {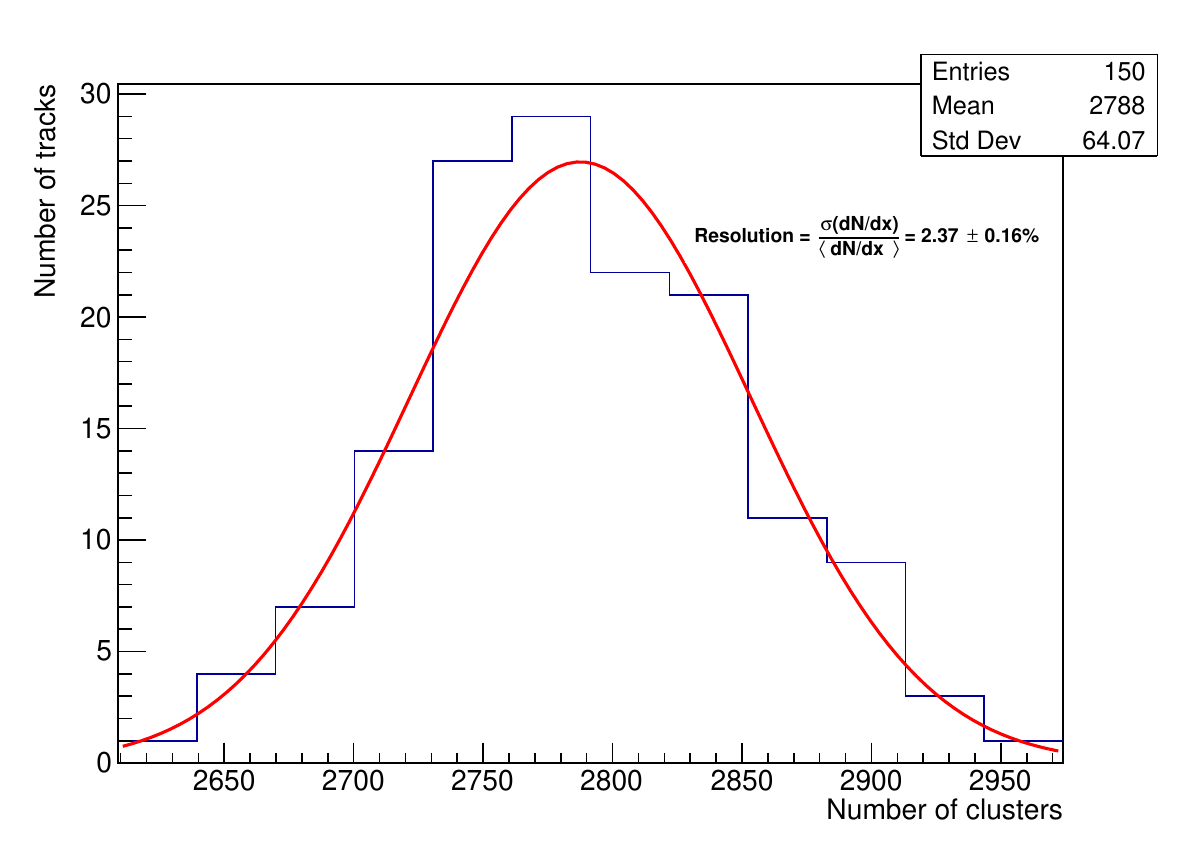}
\includegraphics[width=7.5cm] {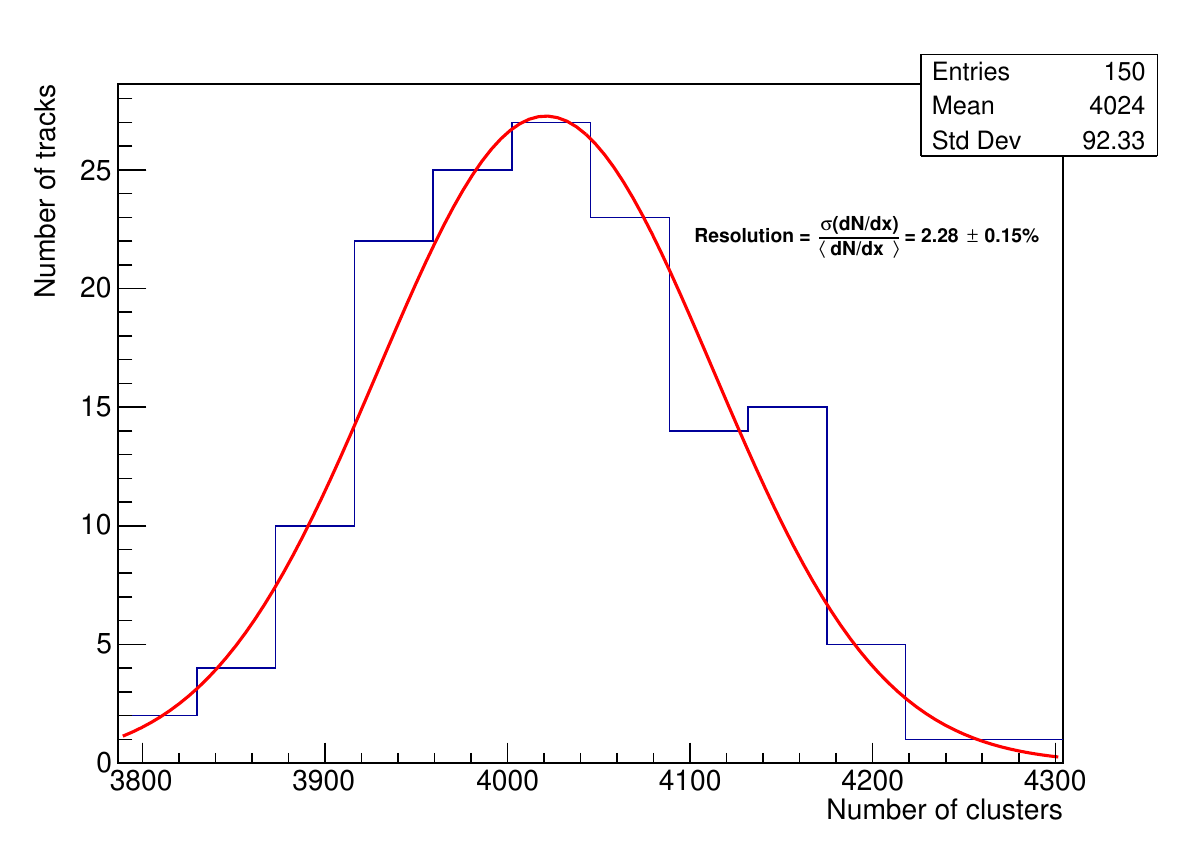}
\caption{Number of ionization clusters measured along 2-meter tracks: after waveform cleaning (left), and after applying both waveform cleaning and recombination and attachment correction (right). The red curves represent the Gaussian fits to the data, with the quoted resolutions including the propagated fit uncertainties.}
\label{Fig:res_2m_Corr}
\end{figure}

\subsubsection{Waveform Cleaning}
To suppress distortions and noise in the raw signal, we introduced a set of cleaning criteria designed to reject anomalous or incomplete waveforms. In particular, we required the time span of each cluster to remain within a physically reasonable range, ensuring consistency with expected signal durations. Tracks exhibiting abnormally wide or noisy waveforms (often indicative of overlapping signals or detector effects) were also excluded. This cleaning procedure effectively reduced the contribution of spurious clusters and improved the overall quality of the waveform reconstruction. As a result, the dN/dx resolution for 2-meter tracks improved from 3.01 $\pm$ 0.20\% to 2.37 $\pm$ 0.16\%, as shown in Figure~\ref{Fig:res_2m_Corr} (left).

\subsubsection{Correction for Recombination and Attachment Effects}
\label{corrClusters}
Ionization electrons drifting toward the readout plane can be partially lost due to recombination and attachment processes. To account for these effects, which reduce the observed number of clusters at longer drift times, a time-dependent correction was applied to the cluster counting on an event-by-event basis.

\begin{figure}[ht]
\centering
\includegraphics[width=\textwidth] {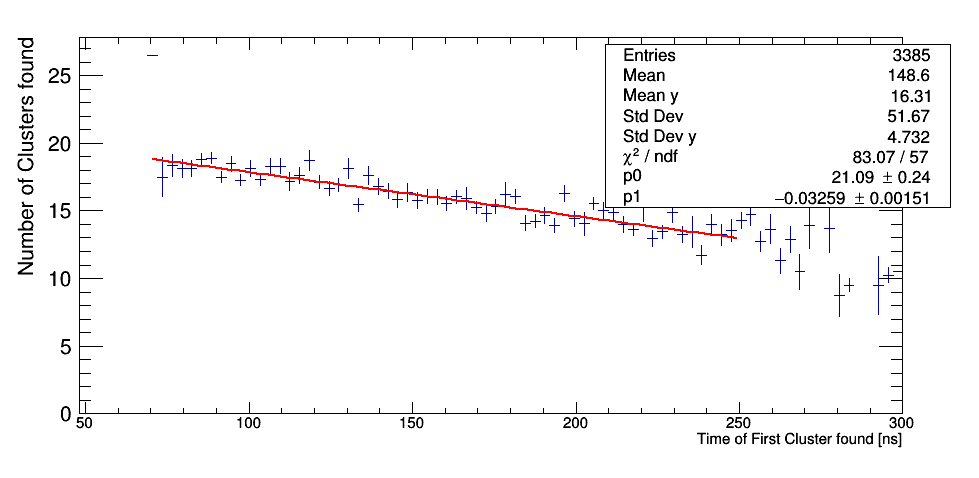}
\caption{The number of clusters as a function of the drift time of the first cluster, for one of the detector configurations. The decreasing trend with drift time reflects the combined effects of electron recombination and attachment.}
\label{Fig:2Dhist}
\end{figure}

A two-dimensional histogram of the number of clusters as a function of the drift time of the first cluster was used to extract a profile along the time axis. To ensure consistency, the histogram must be constructed from events all (coherently) in the same configuration. An example of such a two-dimensional histogram is shown in Figure~\ref{Fig:2Dhist}. This profile was fitted with a first-order polynomial function, and the correction factor (cor) was then computed using the time of the earliest detected cluster in the event, $t_{\text{evt}}$, as follows:

\begin{equation}
 \text{cor} = \frac{a}{a + b \cdot t_{\text{evt}}}
\label{pol1fit}
\end{equation}

where \( a \) and \( b \) are the fit parameters representing the intercept and slope, respectively, characterizing the time dependence of cluster loss.

Then the corrected number of clusters was obtained by:

\begin{equation}
N_{\text{cl, corr}} = N_{\text{cl}} \times \text{cor}
\end{equation}

This correction was derived and applied separately for each detector configuration to account for geometry-dependent variations in drift behavior. As a result, the \( \mathrm{d}N/\mathrm{d}x \) resolution improved further, from 2.37 $\pm$ 0.16\% to 2.28 $\pm$ 0.15\% for 2-meter tracks. The impact of this correction is illustrated in Figure~\ref{Fig:res_2m_Corr} (right).

\section{Conclusions}

The cluster counting technique has proven to be a powerful approach for enhancing particle identification, with both analytical evaluations and simulations demonstrating its efficacy. The development of two key algorithms, DERIV and RTA, has enabled precise detection of electron peaks, with results aligning closely with expected outcomes. The performance of these algorithms was further validated using test beam data under different conditions.

This paper presents a comparative analysis of the resolution performance between the traditional ionization energy loss measurement (\( \mathrm{d}E/\mathrm{d}x \)) and the cluster counting technique (\( \mathrm{d}N/\mathrm{d}x \)) for particle identification. The study confirms that, in its baseline implementation, the cluster counting method achieves a resolution nearly twice as good as the conventional \( \mathrm{d}E/\mathrm{d}x \) approach, in agreement with theoretical expectations.

To further enhance \( \mathrm{d}N/\mathrm{d}x \) performance, two key improvements were introduced: waveform cleaning to reject distorted and noisy signals, and a time-dependent correction to account for electron losses due to recombination and attachment during drift. These refinements reduced the \( \mathrm{d}N/\mathrm{d}x \) resolution from 3.01 $\pm$ 0.20\% to 2.28 $\pm$ 0.15\% for 2-meter tracks, corresponding to an overall improvement by a factor of 2.5 relative to \( \mathrm{d}E/\mathrm{d}x \).

These results highlight the robustness and precision of the cluster counting technique and its potential for future high-precision particle identification applications. Ongoing analysis of the 2023 and 2024 test beam data, particularly in the relativistic rise region, is expected to provide further insights and to guide continued optimization of the method.

\section*{Acknowledgments}
Some individuals have received support by the European Commission with the FEST project, call H2020-MSCA-RISE-2019, contract n. 872901.



\bibliographystyle{JHEP}
 \bibliography{biblio.bib}




\end{document}